\documentclass[iop]{emulateapj}

\usepackage[]{hyperref,epstopdf,amsmath}
\usepackage[]{color,rotating}
\usepackage{ccaption}

\slugcomment{Draft \today}

\shorttitle{Estimating magnetic filling factors}
\shortauthors{V. See et al.}

\begin{document}

\title{Estimating magnetic filling factors from Zeeman-Doppler magnetograms}

\author{Victor See$^{1}$, Sean P. Matt$^{1}$, Colin P. Folsom$^{2,3,4}$, Sudeshna Boro Saikia$^{5}$, \\
Jean-Francois Donati$^{3,4}$, Rim Fares$^{6,7}$, Adam J. Finley$^{1}$,  \'Elodie M. H\'ebrard, \\
Moira M. Jardine$^{8}$, Sandra V. Jeffers$^{9}$,  Lisa T. Lehmann$^{8}$, Stephen C. Marsden$^{7}$, \\
Matthew W. Mengel$^{7}$, Julien Morin$^{10}$, Pascal Petit$^{3,4}$, \\
 Aline A. Vidotto$^{11}$, Ian A. Waite$^{7}$ and the BCool collaboration}
\affil{$^{1}$University of Exeter, Deparment of Physics \& Astronomy, Stocker Road, Devon, Exeter, EX4 4QL, UK\\
$^{2}$Universit\'{e} Grenoble Alpes, CNRS, IPAG, F-38000 Grenoble, France\\
$^{3}$Universit\'{e} de Toulouse, UPS-OMP, IRAP, Toulouse, France\\
$^{4}$CNRS, Institut de Recherche en Astrophysique et Planetologie, 14, avenue Edouard Belin, F-31400 Toulouse, France\\
$^{5}$University of Vienna, Insitute for Astrophysics, Türkenschanzstrasse 17, 1180 Vienna, Austria\\
$^{6}$Physics Department, United Arab Emirates University, P.O. Box 15551, Al-Ain, United Arab Emirates\\
$^{7}$University of Southern Queensland, Centre for Astrophysics, Toowoomba, QLD, 4350, Australia\\
$^{8}$SUPA, School of Physics and Astronomy, University of St Andrews, North Haugh, St Andrews KY16 9SS, UK\\
$^{9}$Universit\"{a}t G\"{o}ttingen, Institut f\"{u}r Astrophysik, Friedrich-Hund-Platz 1, D-37077 G\"{o}ttingen, Germany\\
$^{10}$Laboratoire Univers et Particules de Montpellier, Universit\'{e} ́ de Montpellier, CNRS, F-34095, France\\
$^{11}$School of Physics, Trinity College Dublin, University of Dublin, Dublin-2, Ireland
}
\email{*w.see@exeter.ac.uk}

\begin{abstract}
Low-mass stars are known to have magnetic fields that are believed to be of dynamo origin. Two complementary techniques are principally used to characterise them. Zeeman-Doppler imaging (ZDI) can determine the geometry of the large-scale magnetic field while Zeeman broadening can assess the total unsigned flux including that associated with small-scale structures such as spots. In this work, we study a sample of stars that have been previously mapped with ZDI. We show that the average unsigned magnetic flux follows an activity-rotation relation separating into saturated and unsaturated regimes. We also compare the average photospheric magnetic flux recovered by ZDI, $\langle B_V\rangle$, with that recovered by Zeeman broadening studies, $\langle B_I\rangle$. In line with previous studies, $\langle B_V\rangle$ ranges from a few \% to $\sim$20\% of $\langle B_I\rangle$. We show that a power law relationship between $\langle B_V\rangle$ and $\langle B_I\rangle$ exists and that ZDI recovers a larger fraction of the magnetic flux in more active stars. Using this relation, we improve on previous attempts to estimate filling factors, i.e. the fraction of the stellar surface covered with magnetic field, for stars mapped only with ZDI. Our estimated filling factors follow the well-known activity-rotation relation which is in agreement with filling factors obtained directly from Zeeman broadening studies.  We discuss the possible implications of these results for flux tube expansion above the stellar surface and stellar wind models.
\end{abstract}


\section{Introduction}
\label{sec:Intro}
Over the last two decades, our understanding of stellar magnetism has been enriched by Zeeman-Doppler imaging \citep[ZDI;][]{Donati2009}. This is a tomographic technique that can reconstruct the large-scale photospheric magnetic field topology of low-mass stars from a time-series of high-resolution polarised spectra sampling at least one stellar rotation. \citep{Semel1989,Brown1991,Donati1997,Donati2006}. Repeated observations of individual stars show that their magnetic fields are inherently variable \citep{Morgenthaler2012,Jeffers2014,Saikia2015,Mengel2016,Fares2017,Jeffers2017} and can show regular global polarity reversals similar to those of the Sun \citep{Donati2008TauBoo,Fares2009,Fares2013,Saikia2016,Mengel2016,Jeffers2018,Saikia2018}. Additionally, ensemble studies, that utilise samples consisting of between a handful of stars to nearly 100, have shown that the magnetic properties of low-mass stars depend on fundamental stellar parameters such as mass and rotation \citep{Petit2008,Donati2008,Morin2008,Morin2010,Vidotto2014Trends,See2015,See2016,Folsom2016,Folsom2018}.

Although ZDI is capable of reconstructing the large-scale component of stellar magnetic fields, it is insensitive to small-scale fields, e.g. those associated with magnetic spots. This is because the ZDI technique utilises circularly polarised light (Stokes \textit{V}) which is known to suffer from flux cancellation effects \citep{Morin2010,Johnstone2010,Arzoumanian2011,Lang2014}. A number of authors have studied the link between the large- and small-scale fields by using numerical models \citep{Yadav2015,Lehmann2017,Lehmann2018} or solar magnetograms \citep{Vidotto2016,Vidotto2018}. 

In contrast to ZDI, Zeeman broadening observations make use of unpolarised light (Stokes \textit{I}) that does not suffer from flux cancellation. The disadvantage of using Zeeman broadening is that it is insensitive to magnetic topology. Therefore, ZDI and Zeeman broadening are complementary techniques and both are required to build a holistic picture of stellar magnetism \citep[see][for a summary]{Reiners2012}. Zeeman broadening studies typically express the average unsigned surface field strength, $\langle B_I\rangle$, in terms of a photospheric field strength, $B$, multiplied by a filling factor, $f$, or a combined $Bf  \equiv \langle B_I\rangle$ value\footnote{In this work, we will use $\langle B_I\rangle$ to represent the average unsigned flux from Zeeman broadening studies and $\langle B_V\rangle$ for the average unsigned flux from ZDI studies. We note that these variables have units of Gauss, not Maxwell (or something dimensionally equivalent), despite being called a flux. We direct the interested reader to section 2.1.5 of \citet{Reiners2012} for a more in depth discussion on the terminology of field strengths, fluxes and flux densities.} \citep{Johns-Krull1996,Reiners2007,Phan-Bao2009,ReinersBrowning2009}. Conceptually, $f$ can be thought of as the fraction of the stellar surface filled with magnetic field of strength $B$, with the remaining area, $1-f$, having zero magnetic field. The photospheric field strength, $B$, is thought to be roughly equal to the equipartition field strength \citep{Saar1986}. However, this interpretation is a simplification as studies have shown that Stokes \textit{I} observations can be fit with multiple magnetic components of different field strengths, each with their own associated filling factors, i.e. $\langle B_I\rangle \equiv \sum_i B_{i}f_{i}$ \citep{Johns-Krull2007,Yang2008,Shulyak2014}. Field strengths of up to $\langle B_I\rangle\sim\ 7 {\rm kG}$ have been reported \citep{Shulyak2017} which is well in excess of any surface averaged field strength obtained by ZDI for cool stars. Indeed, comparisons of stars that have been analysed with both Zeeman broadening and ZDI show that the large-scale magnetic flux, to which spectropolarimetry is sensitive, only represents a small fraction of the total magnetic flux \citep{Reiners2009,Morin2010}. Additionally, the rate at which field lines expand with height above the stellar surface is known to affect stellar wind properties \citep{Wang1990,Suzuki2006,Pinto2016}. This rate of expansion is difficult to predict but knowledge of magnetic filling factors can help with this problem \citep{Cranmer2011}.

In this work, we compare the magnetic properties of low-mass stars inferred from Zeeman broadening to those inferred from ZDI. We present a sample of stars that have previously been mapped with ZDI in section \ref{sec:Sample}. In section \ref{sec:MagProperties}, we discuss the magnetic properties of our sample. We present the unsigned magnetic fluxes obtained using ZDI (section \ref{subsec:ZDIProperties}), compare the magnetic field properties of stars that have been observed using both Zeeman broadening and ZDI (section \ref{subsec:ZBvsZDI}) and infer filling factors for our ZDI sample (section \ref{subsec:FillingFactors}). Conclusions are presented in section \ref{sec:Conclusions}.

\section{Sample}
\label{sec:Sample}
In this work, we use a sample of 85 low-mass stars that have had their large-scale photospheric magnetic fields reconstructed with ZDI. A number of these stars have been observed at multiple epochs resulting in a total of 151 magnetic maps in the sample. Many of the ZDI maps come from the efforts of the BCool (Petit et al., in prep) and Toupies \citep{Folsom2016,Folsom2018} collaborations. These stars have a wide range of fundamental parameters with spectral types spanning F, G, K \& M and have rotation periods from fractions of a day to several tens of days. A full list of the stars used can be found in table \ref{tab:Params} along with the average unsigned photospheric flux derived from ZDI, $\langle B_V\rangle$, and references to the original paper for each map. The masses, radii, luminosities and rotation periods of each star are also listed in table \ref{tab:Params}. Unless otherwise noted, these values are taken from the original ZDI publication, \citet{Valenti2005}, \citet{Takeda2007} or from \citet{Vidotto2014Trends} and references therein. In some cases, the bolometric luminosities have been calculated using the $L_{\rm X}/L_{\rm bol}$ and $L_{\rm X}$ values listed in \citet{Vidotto2014Trends}. Rossby numbers are given by the rotation period divided by the convective turnover time.  Convective turnover times are calculated in the manner described by \citet{Cranmer2011} and are a function of effective temperature (see their equation (36)\footnote{\citet{Cranmer2011} state that their equation (36) is valid roughly in the range 3300 K $\lesssim T_{\rm eff} \lesssim$ 7000 K. Although a number of our low-mass stars have effective temperatures below this range, we still use this method to calculate turnover times for these stars. We note that all the stars in our sample with $T_{\rm eff}$ significantly smaller than 3300 K lie in the saturated regime, where magnetic properties do not change significantly. Consequently, the method used to calculate convective turnover times of these stars will not greatly affect our results.}) with an additional weak dependence on the surface gravity for stars with surface gravities smaller than that of the Sun. Finally, $\langle B_I\rangle$ values were available for a subset of the stars in this sample from the literature. These values are listed in table \ref{tab:ZB} along with references for the paper in which they were published.

\section{Magnetic properties}
\label{sec:MagProperties}

\subsection{Zeeman-Doppler imaging}
\label{subsec:ZDIProperties}
ZDI reconstructs the radial, meridional and azimuthal components of the large-scale photospheric stellar magnetic field. Although each map contains a wealth of information, it is common to reduce each map to a set of numerical values that capture the global magnetic field characteristics. In this work, we will use the average unsigned photospheric magnetic flux,  $\langle B_V\rangle$. This is calculated by taking an average of the absolute value of the magnetic field strength over the stellar surface and accounts for all three components of the magnetic field, i.e. radial, meridional and azimuthal.

\begin{figure}
	\begin{center}
	\includegraphics[trim=0cm 1cm 0.5cm 0cm,width=\columnwidth]{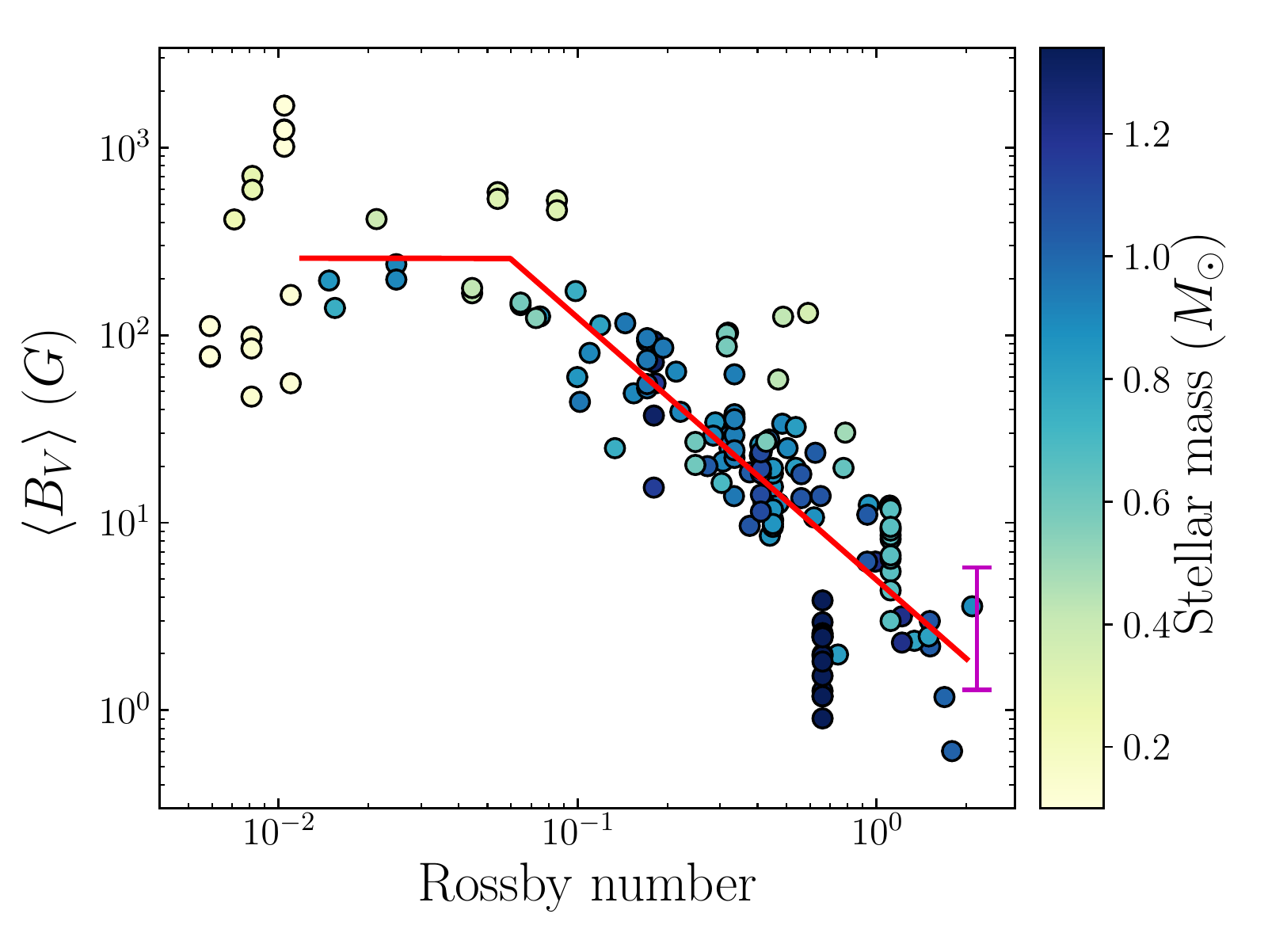}
	\end{center}
	\caption{Average unsigned photospheric magnetic flux obtained from ZDI against Rossby number colour coded by stellar mass. The three parameter fit (solid red line) has a saturated field strength of $\langle B_{V}\rangle_{\rm sat} = 257 \pm 72$G, a critical Rossby number of $\rm Ro_{crit} = 0.06 \pm 0.01$ and an unsaturated regime slope value of $\beta = -1.40 \pm 0.10$. The magenta strut represents the range of $\langle B_{V}\rangle$ values over cycle 24 (the magnetograms used to calculate this range were truncated to $\ell_{\rm max}=5$; see text and \citet{Vidotto2018} for more details).}
	\label{fig:BvsRoss}
\end{figure}

In fig. \ref{fig:BvsRoss}, we plot $\langle B_V\rangle$ against Rossby number. The $\langle B_V\rangle$ values follow the well-known activity-rotation relation shape from studies of other magnetic activity indicators including, but not limited to, X-ray emission \citep{Pizzolato2003,Wright2011,Wright2016,Stelzer2016,Wright2018}, i.e. a roughly constant field strength in the so called ``saturated regime" at small Rossby numbers and a power law relation in the so called ``unsaturated regime" at large Rossby numbers. This is also similar to results found in previous works analysing the relationship between magnetic field properties derived from ZDI and Rossby number \citep{Vidotto2014Trends,See2015,Folsom2016,See2017,Folsom2018}. Additionally, we plot a magenta strut to represent the solar range of $\langle B_V\rangle$ values. This range was calculated using a set of solar magnetograms studied by \citet{Vidotto2018} that cover most of solar cycle 24. Since ZDI only recovers the large-scale magnetic field components, the solar magnetograms were truncated to a spherical harmonic order of $\ell_{\rm max}=5$ to provide a more fair comparison (see \citet{Vidotto2018} for more details). A mean photospheric field strength is derived for each solar magnetogram with the strut representing the range of field strengths seen in these magnetograms.

We perform a three parameter fit to the data of the form

\begin{figure*}
	\begin{center}
	\includegraphics[trim=0cm 1cm 0cm 0cm,width=0.8\textwidth]{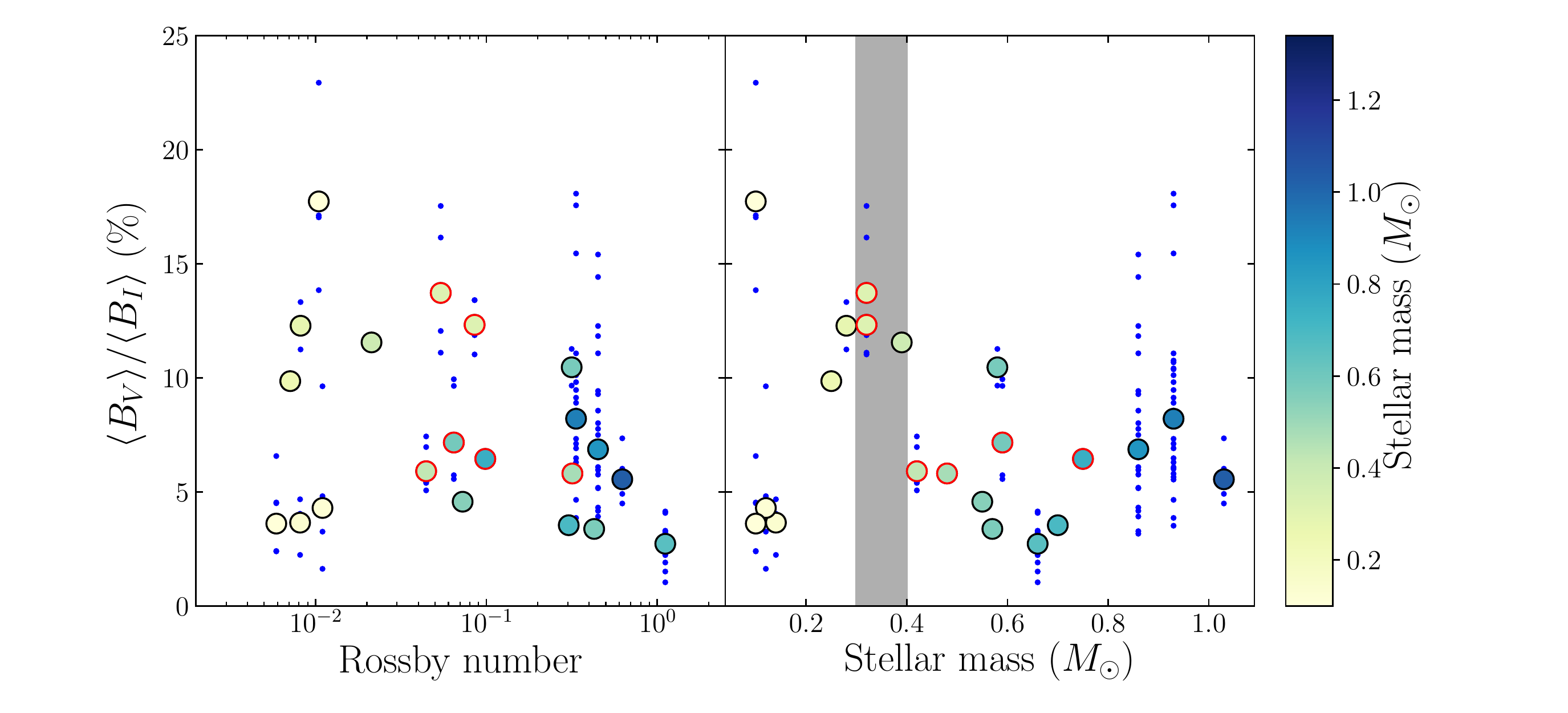}
	\end{center}
	\caption{$\langle B_V\rangle$ as a percentage of $\langle B_I\rangle$ against Rossby number (left) and stellar mass (right). Average values for $\langle B_I\rangle$ and $\langle B_V\rangle$ for each star are shown with large points colour coded by stellar mass. The six stars in the study of \citet{Reiners2009} are outlined in red. All permutations of $\langle B_I\rangle$ and $\langle B_V\rangle$ for each star are shown with small blue points (see text). The shaded region in the right hand panel indicates the transition to full convection ($\sim 0.35M_{\odot}$).}
	\label{fig:Reiners}
\end{figure*}

\begin{equation}
\begin{split}
	\langle B_{V} \rangle & = \langle B_{V}\rangle_{\rm sat} \hspace{23mm} {\rm for\ Ro<Ro_{crit}}  \\
	\langle B_{V} \rangle & = \langle B_{V}\rangle_{\rm sat}\left(\frac{\rm Ro}{\rm Ro_{crit}}\right)^{\beta} \hspace{5mm} {\rm for\ Ro\geq Ro_{crit}} 	
\end{split}	
\label{eq:ActRot}
\end{equation}
where $\langle B_{V}\rangle_{\rm sat}$ is the field strength in the saturated regime, $\rm Ro_{crit}$ is the critical Rossby number dividing the saturated and unsaturated regimes and $\beta$ is the power law index of the unsaturated regime. We find best fit values of $\langle B_{V}\rangle_{\rm sat} = 257 \pm 72$G, $\rm Ro_{crit} = 0.06 \pm 0.01$ and $\beta = -1.40 \pm 0.10$ (shown in fig. \ref{fig:BvsRoss} as a solid red line). The power law slope is relatively well constrained because the majority of data points fall in the unsaturated regime ($\sim$130 maps) and is consistent with the value of $\beta=-1.38 \pm 0.14$ found by \citet{Vidotto2014Trends}. However, $\langle B_{V}\rangle_{\rm sat}$ is less well constrained because there are comparatively few stars in the saturated regime. It is worth noting that the lowest mass stars with the smallest Rossby numbers ($\rm Ro\lesssim 0.012$) have bimodal magnetic fields as previously noted in the literature \citep{Morin2010}. It is clear that these low Rossby number stars are comprised of two sub-groups; one with high field strengths and one with low field strengths. A number of explanations have been proposed for this bimodality \citep{Morin2011,Gastine2013,Kitchatinov2014}. However, there is, as of yet, no consensus and as such, we have excluded these stars from the three parameter fit. Although we have chosen to fit a single saturation level to this data, it is also worth noting that two saturation plateaus may exist if one considers the early-M and mid-M dwarfs separately \citep[see discussion in][]{Vidotto2014Trends}.

An interesting result is the small value we obtain for $\rm Ro_{crit}$. Previous works studying the relationship between different activity indicators and Rossby number typically find critical Rossby numbers that are larger. For example, \citet{Douglas2014} and  \citet{Newton2017} find $\rm Ro_{crit}=0.11^{+0.02}_{-0.03}$ and $\rm Ro_{crit}=0.21 \pm 0.02$ respectively when studying $H\alpha$ emission from different samples, while \citet{Wright2011} find $\rm Ro_{crit}=0.13 \pm 0.02$ when studying X-ray emission. This discrepancy could be due to a number of reasons. For example, we have already noted that the saturation field strength is relatively unconstrained. A larger critical Rossby number could result if the saturation value were lower. Alternatively, differences in the way the convective turnover times, which are notoriously hard to estimate, may contribute to the discrepancy. Lastly, the different $\rm Ro_{crit}$ values may reflect the fact that some of these studies are measuring secondary processes, e.g. X-ray emission, that have non-linear dependencies on the magnetic field strength. As such, it is not obvious that different activity indicators should saturate at the same Rossby number \citep[also see][and references therein]{Jardine1999}.  Further work is required to establish whether the different estimates for $\rm Ro_{crit}$ reflects a real difference in the Rossby number at which large-scale magnetic fields saturate compared to other activity indicators. However, a full comparison of $\rm Ro_{crit}$ values using different activity indicators is beyond the scope of the current work. Finally, we also note that \citet{Reiners2014} suggest that rotation period may be a more relevant parameter compared to Rossby number in the context of magnetic activity. 

\subsection{Zeeman broadening vs ZDI}
\label{subsec:ZBvsZDI}
A growing number of stars have been studied with both ZDI and Zeeman broadening techniques. For each star in our ZDI sample, we search for $\langle B_I \rangle$ values in the literature. This resulted in 21 stars that have at least one $\langle B_V \rangle$ value and one $\langle B_I \rangle$ value. We have listed the $\langle B_I \rangle$ values in table \ref{tab:ZB}. We caution that the $\langle B_I \rangle$ values listed in table \ref{tab:ZB} and the $\langle B_V \rangle$ values listed in table \ref{tab:Params} were not observed simultaneously for any of the stars and this will add some scatter to our plots due to magnetic variability. We also note that these values originate from different authors who have used different models and assumptions that will add an additional level of scatter.

There have been relatively few comparisons between Zeeman broadening observations and ZDI observations in the literature. \citet{Reiners2009} and \citet{Morin2010} compared magnetic field measurements from the two techniques for M stars. A number of key results emerged from these studies. The first is that ZDI only captures a small fraction of the total magnetic flux when compared to Zeeman broadening. The second is that $\langle B_V\rangle /\langle B_I \rangle$ increases by a factor of $\sim 2$ as one crosses the fully convective boundary ($\sim 0.35 M_{\odot}$) from partially to fully convective stars. In fig. \ref{fig:Reiners} we plot $\langle B_V\rangle$ as a percentage of $\langle B_I \rangle$ against Rossby number and stellar mass with the points colour coded by stellar mass. This is similar to the middle panels of figure 2 from \citet{Reiners2009}. Some of the stars have multiple $\langle B_V\rangle$ values, multiple $\langle B_I \rangle$ values or multiple of both. In these cases, we used averaged $\langle B_V \rangle$ or $\langle B_I \rangle$ values. The six stars that were used in the study of \citet{Reiners2009} are outlined in red. Additionally, for each star, we also plot all the combinations of $\langle B_I \rangle$ and $\langle B_V\rangle$ with small blue points. For instance, if a star has $m$ number of $\langle B_I \rangle$ values and $n$ number of $\langle B_V\rangle$ values, it will have a column of $m\times n$ number of blue points around its averaged value in fig. \ref{fig:Reiners}. This visually illustrates the scatter that may exist due to magnetic variability and the fact that the $\langle B_I \rangle$ and $\langle B_V\rangle$ observations were not simultaneous. 

Compared to the studies of \citet{Reiners2009} and \citet{Morin2010}, ours includes a greater number of stars that span a larger range in stellar mass. Similarly to these studies, we find that the reconstructed $\langle B_V\rangle$ value is between a few \% to $\sim 20\%$ of the $\langle B_I \rangle$ value. The second result, that $\langle B_V\rangle /\langle B_I \rangle$ changes by a factor of $\sim$2 across the full convective boundary, still persists but is not as clear in our larger sample. The five stars with masses around or just below the fully convective limit ($M_{\star}\lesssim 0.35M_{\odot}$; EV Lac, GJ 285, V374, Peg EQ Peg A \& EQ Peg B) all have very similar $\langle B_V\rangle /\langle B_I \rangle$ values (around 10\% - 13\%) in line with the results of \citet{Reiners2009} and \citet{Morin2010}. The majority of the partially convective stars have lower average $\langle B_V\rangle /\langle B_I \rangle$ values compared to these 5 fully (or nearly fully) convective stars but there are a few stars worth discussing in more detail. $\epsilon$ Eri ($0.86M_{\odot}$) and $\xi$ Boo A ($0.93M_{\odot}$) both have a large range of $\langle B_V\rangle /\langle B_I \rangle$ values; 3\% to 15\% for $\epsilon$ Eri and 4\% to 18\% for $\xi$ Boo A depending on the combination of $\langle B_V\rangle$ and $\langle B_I \rangle$ values used for each star. The upper values of these $\langle B_V\rangle /\langle B_I \rangle$ ranges are larger than those for the five fully (or nearly fully) convective stars and would seemingly invalidate the conclusion that partially convective stars have lower $\langle B_V\rangle /\langle B_I \rangle$ values. However, this range of $\langle B_V\rangle /\langle B_I \rangle$ values is likely to be an overestimate due to the non-simultaneous observations used to derive the individual $\langle B_V\rangle$ and $\langle B_I \rangle$ values. The true range of possible $\langle B_V\rangle /\langle B_I \rangle$ values is unlikely to be as high or low as suggested by the blue points in fig. \ref{fig:Reiners}. Notably, the average $\langle B_V\rangle /\langle B_I \rangle$ values of 8.2\% for $\xi$ Boo A and 6.9\% for $\epsilon$ Eri are roughly in line with the rest of the partially convective stars. The last star worth briefly discussing is DS Leo ($0.58M_{\odot}$) which has the highest average $\langle B_V\rangle /\langle B_I \rangle$ value of 10.5\% of the partially convective stars. This is comparable to $\langle B_V\rangle /\langle B_I \rangle$ for the five previously discussed fully (or nearly fully) convective stars. Given that it is the only partially convective star with such a high average $\langle B_V\rangle /\langle B_I \rangle$ value, it is unclear if the individual $\langle B_V\rangle$ and $\langle B_I \rangle$ values are discrepent in some way. Simultaneous Stokes \textit{I} and Stokes \textit{V} measurements would be useful to determine whether the $\langle B_V\rangle /\langle B_I \rangle$ value for DS Leo is truly this high. 

At the lowest masses ($\lesssim 0.2M_{\odot}$), we see a wide range $\langle B_V\rangle /\langle B_I \rangle$ values. These stars are a subset of the bimodal $\rm Ro\lesssim0.12$ stars discussed in section \ref{subsec:ZDIProperties}. As noted by \citet{Morin2010} the magnetic fields of these stars are either strong and dipole dominated or comparatively weak and multipolar. These authors also showed that the bimodality is evident when considering $\langle B_V\rangle /\langle B_I \rangle$. WX UMa, which is a strong field dipolar star, has an average $\langle B_V\rangle /\langle B_I \rangle$ value of 18\%. On the other hand, DX Cnc, GJ 1245b and GJ 1156, which are all weak field stars, have average $\langle B_V\rangle /\langle B_I \rangle$ values of $\sim$4\%. Lastly, we note that the upper envelope of $\langle B_V\rangle /\langle B_I \rangle$ points in the left panel of fig. \ref{fig:Reiners} decreases with Rossby number. As noted by \citet{Morin2010}, this may be because all the fully convective stars have small Rossby numbers.

\begin{figure}
	\begin{center}
	\includegraphics[trim=0cm 1cm 0.5cm 0cm,width=\columnwidth]{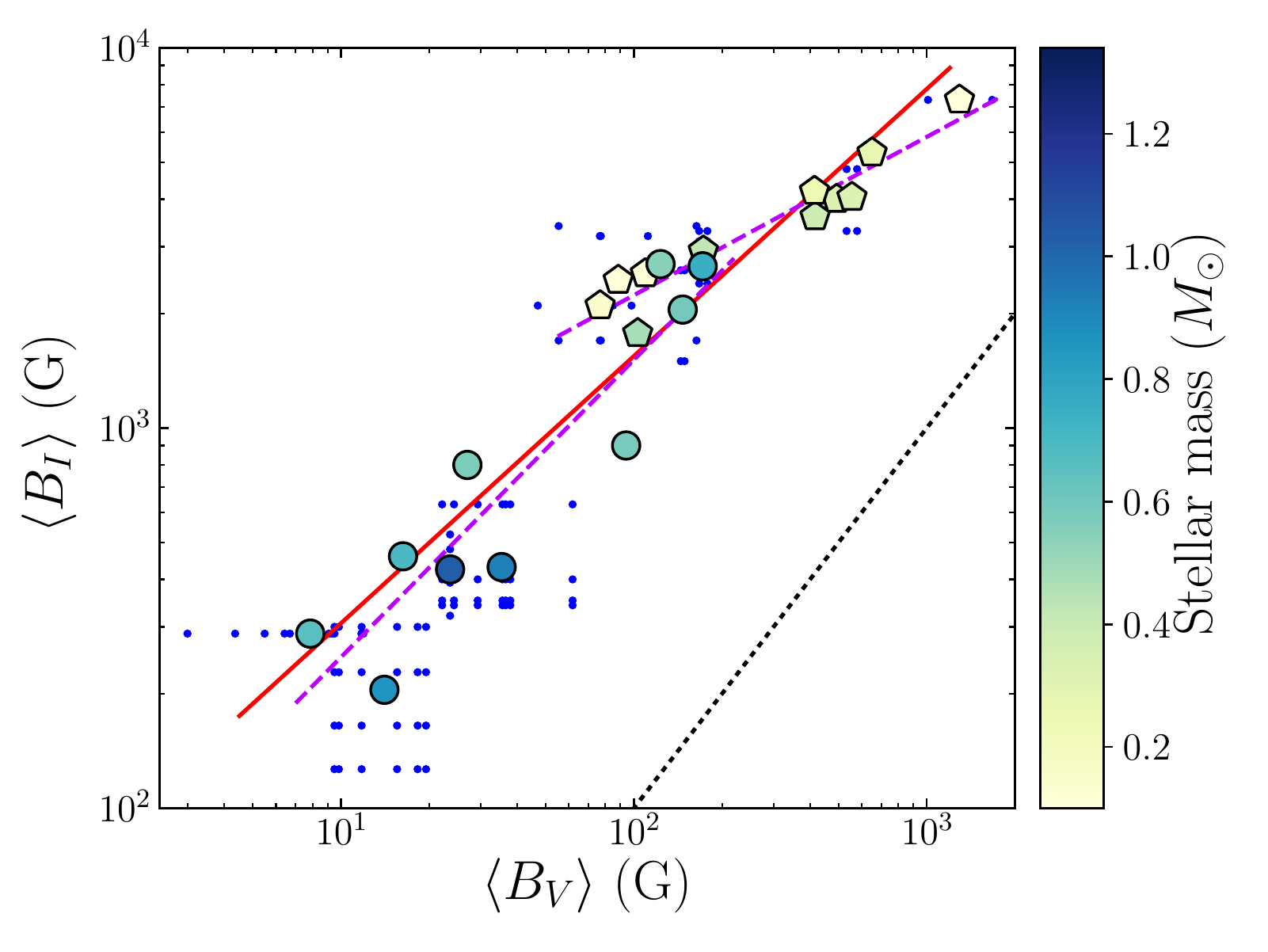}
	\end{center}
	\caption{Average unsigned photospheric magnetic field strengths obtained from Zeeman broadening, $\langle B_I\rangle$, and ZDI, $\langle B_V\rangle$. Symbols have the same meaning as fig. \ref{fig:Reiners}. Stars less massive than 0.5$M_{\odot}$ are shown with pentagons. A best fit line to all the averaged values is shown in red and is given by $\langle B_I\rangle = (61\pm 17)\langle B_V\rangle^{0.70\pm 0.06}$. Fits to stars less massive than 0.5$M_{\odot}$ and more massive than 0.5$M_{\odot}$ are shown with purple dashed lines and are given by $\langle B_I\rangle = (329\pm 79)\langle B_V\rangle^{0.42\pm 0.04}$ and $\langle B_I\rangle = (41\pm 18)\langle B_V\rangle^{0.78\pm 0.12}$ respectively. The black dotted line indicates where $\langle B_I\rangle = \langle B_V\rangle$. }
	\label{fig:BfvsBZDI}
\end{figure}

In fig. \ref{fig:BfvsBZDI}, we plot $\langle B_I \rangle$ directly against $\langle B_V\rangle$. The symbols have the same meanings as in fig. \ref{fig:Reiners} (the small blue points form a cloud around the average point rather than a column in this parameter space). A clear relation between $\langle B_V\rangle $ and $\langle B_I \rangle$ seems to exist. We fit a power law relation to the average points and find that it has the form 

\begin{equation}
	\langle B_I \rangle = (61\pm 17)\langle B_V\rangle^{0.70\pm 0.06},
	\label{eq:BfvsBZDI}
\end{equation}
where $\langle B_I \rangle$ and $\langle B_V \rangle$ are both in units of Gauss. This is shown by a solid red line in fig. \ref{fig:BfvsBZDI}. Again, it is clear that ZDI does not recover all the photospheric flux when comparing the data points to the black dotted line that indicates $\langle B_I \rangle= \langle B_V\rangle$. Taken at face value, equation (\ref{eq:BfvsBZDI}) means that ZDI recovers a larger fraction of the photospheric field for more active stars, i.e. those with larger $\langle B_V\rangle$. Rearranging equation (\ref{eq:BfvsBZDI}), we find $\langle B_V\rangle/\langle B_I\rangle \propto \langle B_V\rangle^{0.29}$. One interpretation is that more active stars may store a smaller fraction of their magnetic energy in small scale structures. \citet{Petit2008} suggested a similar interpretation based on their analysis of the ZDI maps and chromospheric activities of a sample of four stars. This is also backed up by dynamo models that find that the fraction of field in the dipole component goes up for more rapidly rotating, or equivalently, more active, stars (see discussion in section 6.4 of \citet{Brun2017}). If this is true, one might speculate that a higher proportion of the surface magnetic flux is opened up into open flux for more active stars since the open flux is dominated by the large-scale field components \citep[e.g.][]{Jardine2017}. This has implications for calculating stellar angular momentum-loss rates that have been shown to be strongly dependent on the open flux \citep{Reville2015,Pantolmos2017,Finley2017,Finley2018}. On the other hand, this trend may, at least partially, be explained by biases in the ZDI technique since ZDI recovers more small-scale structure for stars with larger $v$sin$i$ \citep{Morin2010}.

An intriguing possibility is that the data points in fig. \ref{fig:BfvsBZDI} can be better fit by two separate power laws. As well as the fit to all the data points given by equation (\ref{eq:BfvsBZDI}), we perform two additional fits to the stars above and below 0.5$M_{\odot}$ separately. These are shown by the dashed purple lines and are given by 

\begin{equation}
	\langle B_I \rangle = (41\pm 18)\langle B_V\rangle^{0.78\pm 0.12}
	\label{eq:BfvsBZDIHighMass}
\end{equation}
and

\begin{equation}
	\langle B_I \rangle = (329\pm 79)\langle B_V\rangle^{0.42\pm 0.04}
	\label{eq:BfvsBZDILowMass}
\end{equation}
respectively. It is apparent, from fig. \ref{fig:BfvsBZDI}, that the two fits have two different power law slopes.  A number of authors have previously discussed a change in the magnetic properties derived from ZDI at 0.5$M_{\odot}$ \citep{Donati2008,Morin2008,Morin2010,Gregory2012,See2015}. For example, \citet{See2015} showed that the energy stored in the toroidal component of the magnetic field increases more steeply as a function of the poloidal magnetic energy for $M_{\star}>0.5M_{\odot}$ stars compared to $M_{\star}<0.5M_{\odot}$ stars (see their fig. 2). This break is very roughly coincident with the mass at which stars become fully convective and may be linked with the change in internal structure. Of course, the two fits are performed on a relatively small number of points and more data will be required to confirm whether the data is truly better fit by two separate power laws. Additionally, we caution that any estimate of $\langle B_I \rangle$ from $\langle B_V\rangle$ using equations (\ref{eq:BfvsBZDI}), (\ref{eq:BfvsBZDIHighMass}) or (\ref{eq:BfvsBZDILowMass}) is only very approximate due to the limited number of data points, the sources of uncertainty discussed previously and intrinsic variability that should be addressed with long-term simultaneous Stokes \textit{I} and Stokes \textit{V} monitoring of these stars.

\subsection{Estimating filling factors}
\label{subsec:FillingFactors}
As discussed in the introduction, $\langle B_I \rangle$ can be interpreted as a fraction of the stellar surface, $f$, filled with magnetic field of strength $B$ \citep{Reiners2012}. \citet{Cranmer2011} showed that the field strength, $B$, is roughly equal to the equipartition field strength, i.e. the field strength that corresponds to balanced magnetic and gas pressures. These authors also showed that the filling factor, $f$, scales with Rossby number following an activity-rotation relation type behaviour. In contrast, ZDI reconstructs magnetic field over the entire stellar surface. There have already been attempts to estimate filling factors for stars based only on ZDI observations. For instance, \citet{Cranmer2017} showed that, by scaling ZDI field strengths by a factor of 7 to account for the flux missed by ZDI, the inferred filling factors are roughly compatible with those found from Zeeman broadening (see their fig. 4). However, these authors note that a more physically motivated correction method could be more appropriate. 

\begin{figure}
	\begin{center}
	\includegraphics[trim=0cm 1cm 0.5cm 0cm,width=\columnwidth]{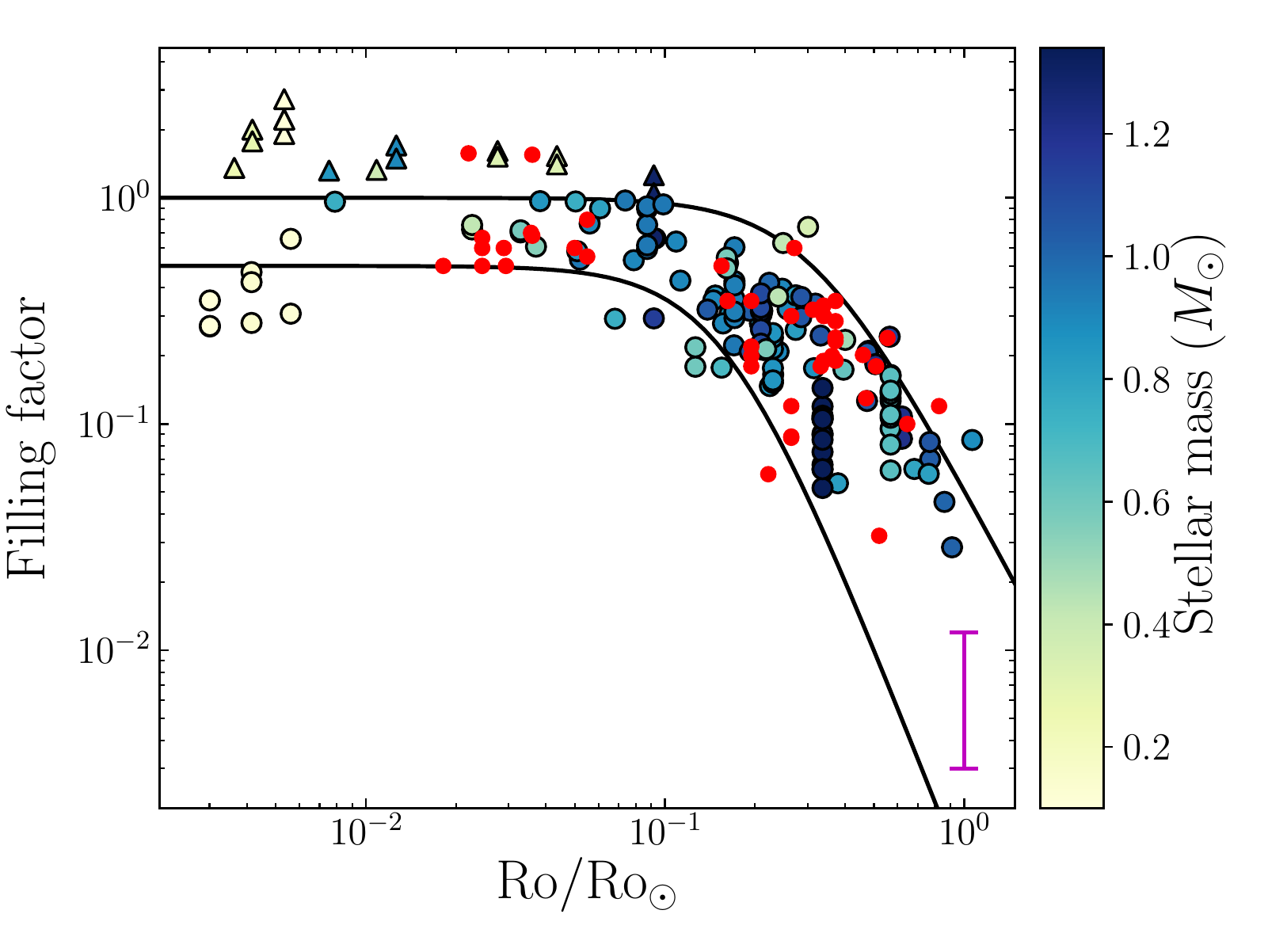}
	\end{center}
	\caption{Filling factor against Rossby number normalised to the solar Rossby number. The data points of \citet{Cranmer2011} are shown in red while their bounding curves are shown in black (see their figure 7b). The range of filling factors exhibited by the Sun, as estimated by \citet{Cranmer2017}, is shown with a magenta strut. The estimated filling factors of the ZDI sample are shown with the coloured points colour coded by stellar mass. Stars with estimated filling factors larger than 1 are shown with triangular points.}
	\label{fig:CS11ff}
\end{figure}

In this section, we will estimate filling factors for our ZDI sample using the following procedure. Using equation (\ref{eq:BfvsBZDI}) and the $\langle B_V\rangle$ value of each ZDI map, we estimate the average surface field strength that Zeeman broadening observations would have retrieved, $\langle B_I \rangle_{\rm est}\ (\equiv B_{\rm est}f_{\rm est})$. We prefer to use equation (\ref{eq:BfvsBZDI}) rather than equations (\ref{eq:BfvsBZDIHighMass}) and (\ref{eq:BfvsBZDILowMass}) since it is not clear whether two separate power law fits are truly justified with the current data. We assume that $B_{\rm est}$ is given by 1.13 times the equipartition field strength following the approach of \citet[][see section 2.1 of their paper for more details]{Cranmer2011}. Using this method, $B_{\rm est}$ scales with the square root of the photospheric density and effective temperature. For our sample, it ranges from $\sim$4kG for the lowest mass stars to $\sim$1kG for the largest. Filling factors are then given by dividing $\langle B_I \rangle_{\rm est}$ by $B_{\rm est}$ and are shown in fig. \ref{fig:CS11ff}. We also show filling factors inferred from Zeeman broadening (red points), bounding envelopes from \citet[][black curves]{Cranmer2011} and the estimated range of the solar filling factor from \citet[][magenta strut]{Cranmer2017} in fig. \ref{fig:CS11ff}. 

Our estimated filling factors broadly follow the activity-rotation relation shape and fall mostly within the two envelopes identified by \citet{Cranmer2011}. On average, more active stars have larger estimated filling factors. This has possible implications for the dynamics of stellar winds. For instance, it is known that the rate at which flux tubes expand can affect stellar wind properties \citep{Wang1990,Suzuki2006,Pinto2016}. The wind carrying flux tubes of more active stars that have larger filling factors are likely to have smaller expansion factors since less expansion is required to fill the circumstellar volume. Care should be taken with this interpretation however because the relevant parameter for stellar winds is the filling factor associated with open flux tubes. In general, this is only a fraction of the total filling factor that we have estimated here. On the Sun, the filling factor of open flux is correlated with the total filling factor over the solar cycle \citep[equation 7 of][]{Cranmer2017} but it is not known whether this relation holds over the course of a cycle on other stars or from star to star.

This method of estimating filling factors from ZDI observations is similar to that of \citet{Cranmer2017}. However, rather than a constant scaling factor of 7 to account for the flux missed by ZDI, we use one that is a function of $\langle B_V\rangle$. Our scaling factors, given by re-arranging equation (\ref{eq:BfvsBZDI}) for $\langle B_I\rangle/\langle B_V\rangle$, range from roughly 60 to 8 for $\langle B_V\rangle$ = 1G to 1kG respectively. This method is likely to be more robust than using a constant scaling factor since it is calibrated using stars that have both ZDI and Zeeman broadening observations. This is reflected in the fact that the majority of the estimated filling factors are roughly consistent with those inferred from Zeeman broadening observations. However, there is still room for improvement. Notably, this method estimates filling factors that are larger than 1 for some of our stars at small Rossby numbers (plotted as triangles in fig. \ref{fig:CS11ff}) which is clearly unphysical. One area of our analysis that could be improved in the future is the assumption that the stellar surface is only covered with either equipartition field or zero field. In reality, the photospheric magnetic field is likely to be highly structured and to have a range of field strengths. Indeed, some observations imply local field strengths that exceed the equipartition field strength \citep{Morin2010,Shulyak2014}. Notably, \citet{Okamoto2018} recently reported an observed field strength of 6.25 kG on the Sun, a value that is roughly four times stronger than equipartition. The fact that we have obtained filling factors larger than 1 could be explained by the lack of super-equipartition field strengths in our calculations. However, it is currently unclear how real magnetic field strengths are distributed on other stars and so we choose to use this simpler model.

\section{Conclusions}
\label{sec:Conclusions}
We have analysed and compared the magnetic properties of low-mass stars derived from two observational techniques. The first is Zeeman-Doppler imaging which is capable of reconstructing the large-scale magnetic field geometry using circularly polarised light (Stokes \textit{V}) but is insensitive to small-scale magnetic field structures such as spots. The second is Zeeman broadening observations which can assess the field down to the smallest scales using unpolarised light (Stokes \textit{I}) but cannot assess field geometry.

In this work, we present the average photospheric unsigned flux from ZDI observations and showed that it follows the well known activity-rotation relation type scaling. There are indications that the critical Rossby number at which the magnetic field strength saturates is smaller than the critical Rossby number from other magnetic activity indicators. In line with previous studies, we confirm that ZDI reconstructs between a few \% and $\sim$20\% of the photospheric magnetic flux and that ZDI seems to recover a smaller percentage of the magnetic flux in partially convective stars than in fully convective stars. At the lowest masses ($\lesssim 0.2M_{\odot}$), there is a large spread in the percentage of magnetic flux that ZDI recovers due to stars with bimodal magnetic fields \citep{Morin2010}.

We find a clear power law relation between the average magnetic fluxes recovered from ZDI and those recovered from Zeeman broadening. There is also a hint that this relationship may be better fit with two separate power laws; one for stars with $M_{\star}<0.5M_{\odot}$ and one for stars with $M_{\star}>0.5M_{\odot}$. However, this suggesting requires more data to confirm, especially for low-mass slow rotators and high-mass fast rotators, which are under-represented in our sample. We use this power law relation to estimate the filling factors for stars that only have ZDI observations. This builds on previous work that has attempted to infer filling factors from ZDI maps \citep{Cranmer2017}. We show that this method produces filling factor estimates that are similar to those obtained from Zeeman broadening studies. These relations allow for a rough assessment of the amount of flux that any given ZDI map may be missing due to flux cancellation effects and will also be helpful for future stellar wind studies. This is because the amount that flux tubes expand above the stellar surface, which depends on the amount of the stellar surface covered in magnetic regions, affects the dynamics of stellar winds \citep{Wang1990}. However, distinguishing the filling factor associated with open flux tubes from the total filling factor remains a challenging task. In the future, our understanding of the relationship between ZDI and Zeeman broadening observations should be improved by the new spectropolarimeter, SPIRou \citep[e.g.][]{Moutou2017}, which will be capable of simultaneous ZDI and Zeeman broadening observations.

\acknowledgments
The authors would like to thank an anonymous referee for their time and useful comments in improving this manuscript and Sam Morrell for useful discussions. VS, SPM and AJF acknowledge funding from the European Research Council (ERC) under the European Unions Horizon 2020 research and innovation programme (grant agreement No 682393 AWESoMeStars). SBS acknowledges funding via the Austrian Space Application Programme (ASAP) of the Austrian Research Promotion Agency (FFG) within ASAP11, the FWF NFN project S11601-N16 and the sub-project S11604-N16. AAV acknowledges funding received from the Irish Research Council Laureate Awards 2017/2018. This work benefitted from discussions within the international team ``The Solar and Stellar Wind Connection: Heating processes and angular momentum loss”, supported by the International Space Science Institute (ISSI).

\bibliographystyle{yahapj}
\bibliography{ZBvsZDI}{}

\begin{thebibliography}{}
\providecommand\natexlab[1]{#1}
\providecommand\JournalTitle[1]{#1}

\bibitem[{{Arzoumanian} {et~al.}(2011){Arzoumanian}, {Jardine}, {Donati},
  {Morin}, \& {Johnstone}}]{Arzoumanian2011}
{Arzoumanian}, D., {Jardine}, M., {Donati}, J.-F., {Morin}, J., \& {Johnstone},
  C. 2011,
  \href{http://dx.doi.org/10.1111/j.1365-2966.2010.17623.x}{\JournalTitle{\mnras},
  410, 2472}

\bibitem[{{Azulay} {et~al.}(2017){Azulay}, {Guirado}, {Marcaide},
  {Mart{\'{\i}}-Vidal}, {Ros}, {Tognelli}, {Jauncey}, {Lestrade}, \&
  {Reynolds}}]{Azulay2017}
{Azulay}, R., {Guirado}, J.~C., {Marcaide}, J.~M., {et~al.} 2017,
  \href{http://dx.doi.org/10.1051/0004-6361/201730641}{\JournalTitle{A\&A},
  607, A10}

\bibitem[{{Boro Saikia} {et~al.}(2015){Boro Saikia}, {Jeffers}, {Petit},
  {Marsden}, {Morin}, \& {Folsom}}]{Saikia2015}
{Boro Saikia}, S., {Jeffers}, S.~V., {Petit}, P., {et~al.} 2015,
  \href{http://dx.doi.org/10.1051/0004-6361/201424096}{\JournalTitle{A\&A},
  573, A17}

\bibitem[{{Boro Saikia} {et~al.}(2016){Boro Saikia}, {Jeffers}, {Morin},
  {Petit}, {Folsom}, {Marsden}, {Donati}, {Cameron}, {Hall}, {Perdelwitz},
  {Reiners}, \& {Vidotto}}]{Saikia2016}
{Boro Saikia}, S., {Jeffers}, S.~V., {Morin}, J., {et~al.} 2016,
  \href{http://dx.doi.org/10.1051/0004-6361/201628262}{\JournalTitle{A\&A},
  594, A29}

\bibitem[{{Boro Saikia} {et~al.}(2018){Boro Saikia}, {Lueftinger}, {Jeffers},
  {Folsom}, {See}, {Petit}, {Marsden}, {Vidotto}, {Morin}, {Reiners}, {Guedel},
  \& {BCool Collaboration}}]{Saikia2018}
{Boro Saikia}, S., {Lueftinger}, T., {Jeffers}, S.~V., {et~al.} 2018,
  \href{http://dx.doi.org/10.1051/0004-6361/201834347}{\JournalTitle{A\&A},
  620, L11}

\bibitem[{{Brown} {et~al.}(1991){Brown}, {Donati}, {Rees}, \&
  {Semel}}]{Brown1991}
{Brown}, S.~F., {Donati}, J.-F., {Rees}, D.~E., \& {Semel}, M. 1991,
  \JournalTitle{A\&A}, 250, 463

\bibitem[{{Brun} \& {Browning}(2017)}]{Brun2017}
{Brun}, A.~S., \& {Browning}, M.~K. 2017,
  \href{http://dx.doi.org/10.1007/s41116-017-0007-8}{\JournalTitle{Living
  Reviews in Solar Physics}, 14, 4}

\bibitem[{{Cranmer}(2017)}]{Cranmer2017}
{Cranmer}, S.~R. 2017,
  \href{http://dx.doi.org/10.3847/1538-4357/aa6f0e}{\JournalTitle{ApJ}, 840,
  114}

\bibitem[{{Cranmer} \& {Saar}(2011)}]{Cranmer2011}
{Cranmer}, S.~R., \& {Saar}, S.~H. 2011,
  \href{http://dx.doi.org/10.1088/0004-637X/741/1/54}{\JournalTitle{ApJ}, 741,
  54}

\bibitem[{{do Nascimento} {et~al.}(2016){do Nascimento}, {Vidotto}, {Petit},
  {Folsom}, {Castro}, {Marsden}, {Morin}, {Porto de Mello}, {Meibom},
  {Jeffers}, {Guinan}, \& {Ribas}}]{Nasciemento2016}
{do Nascimento}, Jr., J.-D., {Vidotto}, A.~A., {Petit}, P., {et~al.} 2016,
  \href{http://dx.doi.org/10.3847/2041-8205/820/1/L15}{\JournalTitle{ApJl},
  820, L15}

\bibitem[{{Donati} \& {Brown}(1997)}]{Donati1997}
{Donati}, J.-F., \& {Brown}, S.~F. 1997, \JournalTitle{A\&A}, 326, 1135

\bibitem[{{Donati} \& {Landstreet}(2009)}]{Donati2009}
{Donati}, J.-F., \& {Landstreet}, J.~D. 2009,
  \href{http://dx.doi.org/10.1146/annurev-astro-082708-101833}{\JournalTitle{ARA\&
  A}, 47, 333}

\bibitem[{{Donati} {et~al.}(2003){Donati}, {Collier Cameron}, {Semel},
  {Hussain}, {Petit}, {Carter}, {Marsden}, {Mengel}, {L{\'o}pez Ariste},
  {Jeffers}, \& {Rees}}]{Donati2003}
{Donati}, J.-F., {Collier Cameron}, A., {Semel}, M., {et~al.} 2003,
  \href{http://dx.doi.org/10.1046/j.1365-2966.2003.07031.x}{\JournalTitle{MNRAS},
  345, 1145}

\bibitem[{{Donati} {et~al.}(2006){Donati}, {Howarth}, {Jardine}, {Petit},
  {Catala}, {Landstreet}, {Bouret}, {Alecian}, {Barnes}, {Forveille},
  {Paletou}, \& {Manset}}]{Donati2006}
{Donati}, J.-F., {Howarth}, I.~D., {Jardine}, M.~M., {et~al.} 2006,
  \href{http://dx.doi.org/10.1111/j.1365-2966.2006.10558.x}{\JournalTitle{MNRAS},
  370, 629}

\bibitem[{{Donati} {et~al.}(2008{\natexlab{a}}){Donati}, {Morin}, {Petit},
  {Delfosse}, {Forveille}, {Auri{\`e}re}, {Cabanac}, {Dintrans}, {Fares},
  {Gastine}, {Jardine}, {Ligni{\`e}res}, {Paletou}, {Ramirez Velez}, \&
  {Th{\'e}ado}}]{Donati2008}
{Donati}, J.-F., {Morin}, J., {Petit}, P., {et~al.} 2008{\natexlab{a}},
  \href{http://dx.doi.org/10.1111/j.1365-2966.2008.13799.x}{\JournalTitle{MNRAS},
  390, 545}

\bibitem[{{Donati} {et~al.}(2008{\natexlab{b}}){Donati}, {Moutou}, {Far{\`e}s},
  {Bohlender}, {Catala}, {Deleuil}, {Shkolnik}, {Collier Cameron}, {Jardine},
  \& {Walker}}]{Donati2008TauBoo}
{Donati}, J.-F., {Moutou}, C., {Far{\`e}s}, R., {et~al.} 2008{\natexlab{b}},
  \href{http://dx.doi.org/10.1111/j.1365-2966.2008.12946.x}{\JournalTitle{MNRAS},
  385, 1179}

\bibitem[{{Douglas} {et~al.}(2014){Douglas}, {Ag{\"u}eros}, {Covey}, {Bowsher},
  {Bochanski}, {Cargile}, {Kraus}, {Law}, {Lemonias}, {Arce}, {Fierroz}, \&
  {Kundert}}]{Douglas2014}
{Douglas}, S.~T., {Ag{\"u}eros}, M.~A., {Covey}, K.~R., {et~al.} 2014,
  \href{http://dx.doi.org/10.1088/0004-637X/795/2/161}{\JournalTitle{ApJ}, 795,
  161}

\bibitem[{{Fares} {et~al.}(2013){Fares}, {Moutou}, {Donati}, {Catala},
  {Shkolnik}, {Jardine}, {Cameron}, \& {Deleuil}}]{Fares2013}
{Fares}, R., {Moutou}, C., {Donati}, J.-F., {et~al.} 2013,
  \href{http://dx.doi.org/10.1093/mnras/stt1386}{\JournalTitle{MNRAS}, 435,
  1451}

\bibitem[{{Fares} {et~al.}(2009){Fares}, {Donati}, {Moutou}, {Bohlender},
  {Catala}, {Deleuil}, {Shkolnik}, {Collier Cameron}, {Jardine}, \&
  {Walker}}]{Fares2009}
{Fares}, R., {Donati}, J.-F., {Moutou}, C., {et~al.} 2009,
  \href{http://dx.doi.org/10.1111/j.1365-2966.2009.15303.x}{\JournalTitle{MNRAS},
  398, 1383}

\bibitem[{{Fares} {et~al.}(2010){Fares}, {Donati}, {Moutou}, {Jardine},
  {Grie{\ss}meier}, {Zarka}, {Shkolnik}, {Bohlender}, {Catala}, \& {Collier
  Cameron}}]{Fares2010}
---. 2010,
  \href{http://dx.doi.org/10.1111/j.1365-2966.2010.16715.x}{\JournalTitle{MNRAS},
  406, 409}

\bibitem[{{Fares} {et~al.}(2012){Fares}, {Donati}, {Moutou}, {Jardine},
  {Cameron}, {Lanza}, {Bohlender}, {Dieters}, {Mart{\'{\i}}nez Fiorenzano},
  {Maggio}, {Pagano}, \& {Shkolnik}}]{Fares2012}
---. 2012,
  \href{http://dx.doi.org/10.1111/j.1365-2966.2012.20780.x}{\JournalTitle{MNRAS},
  423, 1006}

\bibitem[{{Fares} {et~al.}(2017){Fares}, {Bourrier}, {Vidotto}, {Moutou},
  {Jardine}, {Zarka}, {Helling}, {Lecavelier des Etangs}, {Llama}, {Louden},
  {Wheatley}, \& {Ehrenreich}}]{Fares2017}
{Fares}, R., {Bourrier}, V., {Vidotto}, A.~A., {et~al.} 2017,
  \href{http://dx.doi.org/10.1093/mnras/stx1581}{\JournalTitle{MNRAS}, 471,
  1246}

\bibitem[{{Fernandes} {et~al.}(1998){Fernandes}, {Lebreton}, {Baglin}, \&
  {Morel}}]{Fernandes1998}
{Fernandes}, J., {Lebreton}, Y., {Baglin}, A., \& {Morel}, P. 1998,
  \JournalTitle{A\&A}, 338, 455

\bibitem[{{Finley} \& {Matt}(2017)}]{Finley2017}
{Finley}, A.~J., \& {Matt}, S.~P. 2017,
  \href{http://dx.doi.org/10.3847/1538-4357/aa7fb9}{\JournalTitle{ApJ}, 845,
  46}

\bibitem[{{Finley} \& {Matt}(2018)}]{Finley2018}
---. 2018,
  \href{http://dx.doi.org/10.3847/1538-4357/aaaab5}{\JournalTitle{ApJ}, 854,
  78}

\bibitem[{{Folsom} {et~al.}(2016){Folsom}, {Petit}, {Bouvier}, {L{\`e}bre},
  {Amard}, {Palacios}, {Morin}, {Donati}, {Jeffers}, {Marsden}, \&
  {Vidotto}}]{Folsom2016}
{Folsom}, C.~P., {Petit}, P., {Bouvier}, J., {et~al.} 2016,
  \href{http://dx.doi.org/10.1093/mnras/stv2924}{\JournalTitle{MNRAS}, 457,
  580}

\bibitem[{{Folsom} {et~al.}(2018{\natexlab{a}}){Folsom}, {Fossati}, {Wood},
  {Sreejith}, {Cubillos}, {Vidotto}, {Alecian}, {Girish}, {Lichtenegger},
  {Murthy}, {Petit}, \& {Valyavin}}]{Folsom2018Single}
{Folsom}, C.~P., {Fossati}, L., {Wood}, B.~E., {et~al.} 2018{\natexlab{a}},
  \JournalTitle{ArXiv e-prints},
  \href{http://arxiv.org/abs/1808.00406}{{\sffamily arXiv:1808.00406
  [astro-ph.SR]}}

\bibitem[{{Folsom} {et~al.}(2018{\natexlab{b}}){Folsom}, {Bouvier}, {Petit},
  {L{\`e}bre}, {Amard}, {Palacios}, {Morin}, {Donati}, \&
  {Vidotto}}]{Folsom2018}
{Folsom}, C.~P., {Bouvier}, J., {Petit}, P., {et~al.} 2018{\natexlab{b}},
  \href{http://dx.doi.org/10.1093/mnras/stx3021}{\JournalTitle{MNRAS}, 474,
  4956}

\bibitem[{{Gastine} {et~al.}(2013){Gastine}, {Morin}, {Duarte}, {Reiners},
  {Christensen}, \& {Wicht}}]{Gastine2013}
{Gastine}, T., {Morin}, J., {Duarte}, L., {et~al.} 2013,
  \href{http://dx.doi.org/10.1051/0004-6361/201220317}{\JournalTitle{A\&A},
  549, L5}

\bibitem[{{Gillon} {et~al.}(2017){Gillon}, {Demory}, {Van Grootel}, {Motalebi},
  {Lovis}, {Cameron}, {Charbonneau}, {Latham}, {Molinari}, {Pepe},
  {S{\'e}gransan}, {Sasselov}, {Udry}, {Mayor}, {Micela}, {Piotto}, \&
  {Sozzetti}}]{Gillon2017}
{Gillon}, M., {Demory}, B.-O., {Van Grootel}, V., {et~al.} 2017,
  \href{http://dx.doi.org/10.1038/s41550-017-0056}{\JournalTitle{Nature
  Astronomy}, 1, 0056}

\bibitem[{{Gregory} {et~al.}(2012){Gregory}, {Donati}, {Morin}, {Hussain},
  {Mayne}, {Hillenbrand}, \& {Jardine}}]{Gregory2012}
{Gregory}, S.~G., {Donati}, J.-F., {Morin}, J., {et~al.} 2012,
  \href{http://dx.doi.org/10.1088/0004-637X/755/2/97}{\JournalTitle{ApJ}, 755,
  97}

\bibitem[{{Guirado} {et~al.}(2011){Guirado}, {Marcaide}, {Mart{\'{\i}}-Vidal},
  {Le Bouquin}, {Close}, {Cotton}, \& {Montalb{\'a}n}}]{Guirado2011}
{Guirado}, J.~C., {Marcaide}, J.~M., {Mart{\'{\i}}-Vidal}, I., {et~al.} 2011,
  \href{http://dx.doi.org/10.1051/0004-6361/201117426}{\JournalTitle{A\&A},
  533, A106}

\bibitem[{{H{\'e}brard} {et~al.}(2016){H{\'e}brard}, {Donati}, {Delfosse},
  {Morin}, {Moutou}, \& {Boisse}}]{Hebrard2016}
{H{\'e}brard}, {\'E}.~M., {Donati}, J.-F., {Delfosse}, X., {et~al.} 2016,
  \href{http://dx.doi.org/10.1093/mnras/stw1346}{\JournalTitle{MNRAS}, 461,
  1465}

\bibitem[{{Jardine} \& {Unruh}(1999)}]{Jardine1999}
{Jardine}, M., \& {Unruh}, Y.~C. 1999, \JournalTitle{A\&A}, 346, 883

\bibitem[{{Jardine} {et~al.}(2017){Jardine}, {Vidotto}, \& {See}}]{Jardine2017}
{Jardine}, M., {Vidotto}, A.~A., \& {See}, V. 2017,
  \href{http://dx.doi.org/10.1093/mnrasl/slw206}{\JournalTitle{MNRAS}, 465,
  L25}

\bibitem[{{Jeffers} {et~al.}(2017){Jeffers}, {Boro Saikia}, {Barnes}, {Petit},
  {Marsden}, {Jardine}, {Vidotto}, \& {BCool Collaboration}}]{Jeffers2017}
{Jeffers}, S.~V., {Boro Saikia}, S., {Barnes}, J.~R., {et~al.} 2017,
  \href{http://dx.doi.org/10.1093/mnrasl/slx097}{\JournalTitle{MNRAS}, 471,
  L96}

\bibitem[{{Jeffers} {et~al.}(2014){Jeffers}, {Petit}, {Marsden}, {Morin},
  {Donati}, \& {Folsom}}]{Jeffers2014}
{Jeffers}, S.~V., {Petit}, P., {Marsden}, S.~C., {et~al.} 2014,
  \href{http://dx.doi.org/10.1051/0004-6361/201423725}{\JournalTitle{A\&A},
  569, A79}

\bibitem[{{Jeffers} {et~al.}(2018){Jeffers}, {Mengel}, {Moutou}, {Marsden},
  {Barnes}, {Jardine}, {Petit}, {Schmitt}, {See}, {Vidotto}, \& {BCool
  Collaboration}}]{Jeffers2018}
{Jeffers}, S.~V., {Mengel}, M., {Moutou}, C., {et~al.} 2018,
  \href{http://dx.doi.org/10.1093/mnras/sty1717}{\JournalTitle{MNRAS}, 479,
  5266}

\bibitem[{{Johns-Krull}(2007)}]{Johns-Krull2007}
{Johns-Krull}, C.~M. 2007,
  \href{http://dx.doi.org/10.1086/519017}{\JournalTitle{ApJ}, 664, 975}

\bibitem[{{Johns-Krull} \& {Valenti}(1996)}]{Johns-Krull1996}
{Johns-Krull}, C.~M., \& {Valenti}, J.~A. 1996,
  \href{http://dx.doi.org/10.1086/309954}{\JournalTitle{ApJL}, 459, L95}

\bibitem[{{Johnstone} {et~al.}(2010){Johnstone}, {Jardine}, \&
  {Mackay}}]{Johnstone2010}
{Johnstone}, C., {Jardine}, M., \& {Mackay}, D.~H. 2010,
  \href{http://dx.doi.org/10.1111/j.1365-2966.2010.16298.x}{\JournalTitle{MNRAS},
  404, 101}

\bibitem[{{Kitchatinov} {et~al.}(2014){Kitchatinov}, {Moss}, \&
  {Sokoloff}}]{Kitchatinov2014}
{Kitchatinov}, L.~L., {Moss}, D., \& {Sokoloff}, D. 2014,
  \href{http://dx.doi.org/10.1093/mnrasl/slu041}{\JournalTitle{MNRAS}, 442, L1}

\bibitem[{{Lang} {et~al.}(2014){Lang}, {Jardine}, {Morin}, {Donati}, {Jeffers},
  {Vidotto}, \& {Fares}}]{Lang2014}
{Lang}, P., {Jardine}, M., {Morin}, J., {et~al.} 2014,
  \href{http://dx.doi.org/10.1093/mnras/stu091}{\JournalTitle{MNRAS}, 439,
  2122}

\bibitem[{{Lehmann} {et~al.}(2018){Lehmann}, {Jardine}, {Mackay}, \&
  {Vidotto}}]{Lehmann2018}
{Lehmann}, L.~T., {Jardine}, M.~M., {Mackay}, D.~H., \& {Vidotto}, A.~A. 2018,
  \href{http://dx.doi.org/10.1093/mnras/sty1230}{\JournalTitle{\mnras}, 478,
  4390}

\bibitem[{{Lehmann} {et~al.}(2017){Lehmann}, {Jardine}, {Vidotto}, {Mackay},
  {See}, {Donati}, {Folsom}, {Jeffers}, {Marsden}, {Morin}, \&
  {Petit}}]{Lehmann2017}
{Lehmann}, L.~T., {Jardine}, M.~M., {Vidotto}, A.~A., {et~al.} 2017,
  \href{http://dx.doi.org/10.1093/mnrasl/slw225}{\JournalTitle{MNRAS}, 466,
  L24}

\bibitem[{{Marcy} \& {Basri}(1989)}]{Marcy1989}
{Marcy}, G.~W., \& {Basri}, G. 1989,
  \href{http://dx.doi.org/10.1086/167921}{\JournalTitle{\apj}, 345, 480}

\bibitem[{{Marsden} {et~al.}(2011){Marsden}, {Jardine}, {Ram{\'{\i}}rez
  V{\'e}lez}, {Alecian}, {Brown}, {Carter}, {Donati}, {Dunstone}, {Hart},
  {Semel}, \& {Waite}}]{Marsden2011}
{Marsden}, S.~C., {Jardine}, M.~M., {Ram{\'{\i}}rez V{\'e}lez}, J.~C., {et~al.}
  2011,
  \href{http://dx.doi.org/10.1111/j.1365-2966.2011.18367.x}{\JournalTitle{MNRAS},
  413, 1922}

\bibitem[{{Mengel} {et~al.}(2016){Mengel}, {Fares}, {Marsden}, {Carter},
  {Jeffers}, {Petit}, {Donati}, {Folsom}, \& {BCool
  Collaboration}}]{Mengel2016}
{Mengel}, M.~W., {Fares}, R., {Marsden}, S.~C., {et~al.} 2016,
  \href{http://dx.doi.org/10.1093/mnras/stw828}{\JournalTitle{MNRAS}, 459,
  4325}

\bibitem[{{Montesinos} \& {Jordan}(1993)}]{Montesinos1993}
{Montesinos}, B., \& {Jordan}, C. 1993,
  \href{http://dx.doi.org/10.1093/mnras/264.4.900}{\JournalTitle{MNRAS}, 264,
  900}

\bibitem[{{Morgenthaler} {et~al.}(2012){Morgenthaler}, {Petit}, {Saar},
  {Solanki}, {Morin}, {Marsden}, {Auri{\`e}re}, {Dintrans}, {Fares}, {Gastine},
  {Lanoux}, {Ligni{\`e}res}, {Paletou}, {Ram{\'{\i}}rez V{\'e}lez},
  {Th{\'e}ado}, \& {Van Grootel}}]{Morgenthaler2012}
{Morgenthaler}, A., {Petit}, P., {Saar}, S., {et~al.} 2012,
  \href{http://dx.doi.org/10.1051/0004-6361/201118139}{\JournalTitle{A\& A},
  540, A138}

\bibitem[{{Morin} {et~al.}(2010){Morin}, {Donati}, {Petit}, {Delfosse},
  {Forveille}, \& {Jardine}}]{Morin2010}
{Morin}, J., {Donati}, J.-F., {Petit}, P., {et~al.} 2010,
  \href{http://dx.doi.org/10.1111/j.1365-2966.2010.17101.x}{\JournalTitle{MNRAS},
  407, 2269}

\bibitem[{{Morin} {et~al.}(2011){Morin}, {Dormy}, {Schrinner}, \&
  {Donati}}]{Morin2011}
{Morin}, J., {Dormy}, E., {Schrinner}, M., \& {Donati}, J.-F. 2011,
  \href{http://dx.doi.org/10.1111/j.1745-3933.2011.01159.x}{\JournalTitle{MNRAS},
  418, L133}

\bibitem[{{Morin} {et~al.}(2008{\natexlab{a}}){Morin}, {Donati}, {Petit},
  {Delfosse}, {Forveille}, {Albert}, {Auri{\`e}re}, {Cabanac}, {Dintrans},
  {Fares}, {Gastine}, {Jardine}, {Ligni{\`e}res}, {Paletou}, {Ramirez Velez},
  \& {Th{\'e}ado}}]{Morin2008}
{Morin}, J., {Donati}, J.-F., {Petit}, P., {et~al.} 2008{\natexlab{a}},
  \href{http://dx.doi.org/10.1111/j.1365-2966.2008.13809.x}{\JournalTitle{MNRAS},
  390, 567}

\bibitem[{{Morin} {et~al.}(2008{\natexlab{b}}){Morin}, {Donati}, {Forveille},
  {Delfosse}, {Dobler}, {Petit}, {Jardine}, {Collier Cameron}, {Albert},
  {Manset}, {Dintrans}, {Chabrier}, \& {Valenti}}]{Morin2008V374Peg}
{Morin}, J., {Donati}, J.-F., {Forveille}, T., {et~al.} 2008{\natexlab{b}},
  \href{http://dx.doi.org/10.1111/j.1365-2966.2007.12709.x}{\JournalTitle{MNRAS},
  384, 77}

\bibitem[{{Moutou} {et~al.}(2017){Moutou}, {H{\'e}brard}, {Morin}, {Malo},
  {Fouqu{\'e}}, {Torres-Rivas}, {Martioli}, {Delfosse}, {Artigau}, \&
  {Doyon}}]{Moutou2017}
{Moutou}, C., {H{\'e}brard}, E.~M., {Morin}, J., {et~al.} 2017,
  \href{http://dx.doi.org/10.1093/mnras/stx2306}{\JournalTitle{MNRAS}, 472,
  4563}

\bibitem[{{Newton} {et~al.}(2017){Newton}, {Irwin}, {Charbonneau}, {Berlind},
  {Calkins}, \& {Mink}}]{Newton2017}
{Newton}, E.~R., {Irwin}, J., {Charbonneau}, D., {et~al.} 2017,
  \href{http://dx.doi.org/10.3847/1538-4357/834/1/85}{\JournalTitle{ApJ}, 834,
  85}

\bibitem[{{Okamoto} \& {Sakurai}(2018)}]{Okamoto2018}
{Okamoto}, T.~J., \& {Sakurai}, T. 2018,
  \href{http://dx.doi.org/10.3847/2041-8213/aaa3d8}{\JournalTitle{ApJl}, 852,
  L16}

\bibitem[{{Pantolmos} \& {Matt}(2017)}]{Pantolmos2017}
{Pantolmos}, G., \& {Matt}, S.~P. 2017,
  \href{http://dx.doi.org/10.3847/1538-4357/aa9061}{\JournalTitle{ApJ}, 849,
  83}

\bibitem[{{Petit} {et~al.}(2008){Petit}, {Dintrans}, {Solanki}, {Donati},
  {Auri{\`e}re}, {Ligni{\`e}res}, {Morin}, {Paletou}, {Ramirez Velez},
  {Catala}, \& {Fares}}]{Petit2008}
{Petit}, P., {Dintrans}, B., {Solanki}, S.~K., {et~al.} 2008,
  \href{http://dx.doi.org/10.1111/j.1365-2966.2008.13411.x}{\JournalTitle{MNRAS},
  388, 80}

\bibitem[{{Phan-Bao} {et~al.}(2009){Phan-Bao}, {Lim}, {Donati}, {Johns-Krull},
  \& {Mart{\'{\i}}n}}]{Phan-Bao2009}
{Phan-Bao}, N., {Lim}, J., {Donati}, J.-F., {Johns-Krull}, C.~M., \&
  {Mart{\'{\i}}n}, E.~L. 2009,
  \href{http://dx.doi.org/10.1088/0004-637X/704/2/1721}{\JournalTitle{ApJ},
  704, 1721}

\bibitem[{{Pinto} {et~al.}(2016){Pinto}, {Brun}, \& {Rouillard}}]{Pinto2016}
{Pinto}, R.~F., {Brun}, A.~S., \& {Rouillard}, A.~P. 2016,
  \href{http://dx.doi.org/10.1051/0004-6361/201628599}{\JournalTitle{A\&A},
  592, A65}

\bibitem[{{Pizzolato} {et~al.}(2003){Pizzolato}, {Maggio}, {Micela},
  {Sciortino}, \& {Ventura}}]{Pizzolato2003}
{Pizzolato}, N., {Maggio}, A., {Micela}, G., {Sciortino}, S., \& {Ventura}, P.
  2003, \href{http://dx.doi.org/10.1051/0004-6361:20021560}{\JournalTitle{aap},
  397, 147}

\bibitem[{{Reiners}(2012)}]{Reiners2012}
{Reiners}, A. 2012,
  \href{http://dx.doi.org/10.12942/lrsp-2012-1}{\JournalTitle{Living Reviews in
  Solar Physics}, 9, 1}

\bibitem[{{Reiners} \& {Basri}(2007)}]{Reiners2007}
{Reiners}, A., \& {Basri}, G. 2007,
  \href{http://dx.doi.org/10.1086/510304}{\JournalTitle{ApJ}, 656, 1121}

\bibitem[{{Reiners} \& {Basri}(2009)}]{Reiners2009}
---. 2009,
  \href{http://dx.doi.org/10.1051/0004-6361:200811450}{\JournalTitle{A\&A},
  496, 787}

\bibitem[{{Reiners} {et~al.}(2009{\natexlab{a}}){Reiners}, {Basri}, \&
  {Browning}}]{ReinersBrowning2009}
{Reiners}, A., {Basri}, G., \& {Browning}, M. 2009{\natexlab{a}},
  \href{http://dx.doi.org/10.1088/0004-637X/692/1/538}{\JournalTitle{ApJ}, 692,
  538}

\bibitem[{{Reiners} {et~al.}(2009{\natexlab{b}}){Reiners}, {Basri}, \&
  {Browning}}]{Reiners2009FluxSat}
---. 2009{\natexlab{b}},
  \href{http://dx.doi.org/10.1088/0004-637X/692/1/538}{\JournalTitle{ApJ}, 692,
  538}

\bibitem[{{Reiners} {et~al.}(2014){Reiners}, {Sch{\"u}ssler}, \&
  {Passegger}}]{Reiners2014}
{Reiners}, A., {Sch{\"u}ssler}, M., \& {Passegger}, V.~M. 2014,
  \href{http://dx.doi.org/10.1088/0004-637X/794/2/144}{\JournalTitle{ApJ}, 794,
  144}

\bibitem[{{R{\'e}ville} {et~al.}(2015){R{\'e}ville}, {Brun}, {Matt},
  {Strugarek}, \& {Pinto}}]{Reville2015}
{R{\'e}ville}, V., {Brun}, A.~S., {Matt}, S.~P., {Strugarek}, A., \& {Pinto},
  R.~F. 2015,
  \href{http://dx.doi.org/10.1088/0004-637X/798/2/116}{\JournalTitle{ApJ}, 798,
  116}

\bibitem[{{Saar}(1994)}]{Saar1994}
{Saar}, S.~H. 1994, in IAU Symposium, Vol. 154, Infrared Solar Physics, ed.
  D.~M. {Rabin}, J.~T. {Jefferies}, \& C.~{Lindsey}, 493

\bibitem[{{Saar}(2001)}]{Saar2001}
{Saar}, S.~H. 2001, in Astronomical Society of the Pacific Conference Series,
  Vol. 223, 11th Cambridge Workshop on Cool Stars, Stellar Systems and the Sun,
  ed. R.~J. {Garcia Lopez}, R.~{Rebolo}, \& M.~R. {Zapaterio Osorio}, 292

\bibitem[{{Saar} \& {Linsky}(1986)}]{Saar1986}
{Saar}, S.~H., \& {Linsky}, J.~L. 1986,
  \href{http://dx.doi.org/10.1016/0273-1177(86)90444-8}{\JournalTitle{Advances
  in Space Research}, 6, 235}

\bibitem[{{See} {et~al.}(2015){See}, {Jardine}, {Vidotto}, {Donati}, {Folsom},
  {Boro Saikia}, {Bouvier}, {Fares}, {Gregory}, {Hussain}, {Jeffers},
  {Marsden}, {Morin}, {Moutou}, {do Nascimento}, {Petit}, {Ros{\'e}n}, \&
  {Waite}}]{See2015}
{See}, V., {Jardine}, M., {Vidotto}, A.~A., {et~al.} 2015,
  \href{http://dx.doi.org/10.1093/mnras/stv1925}{\JournalTitle{MNRAS}, 453,
  4301}

\bibitem[{{See} {et~al.}(2016){See}, {Jardine}, {Vidotto}, {Donati}, {Boro
  Saikia}, {Bouvier}, {Fares}, {Folsom}, {Gregory}, {Hussain}, {Jeffers},
  {Marsden}, {Morin}, {Moutou}, {do Nascimento}, {Petit}, \& {Waite}}]{See2016}
---. 2016, \href{http://dx.doi.org/10.1093/mnras/stw2010}{\JournalTitle{MNRAS},
  462, 4442}

\bibitem[{{See} {et~al.}(2017){See}, {Jardine}, {Vidotto}, {Donati}, {Boro
  Saikia}, {Fares}, {Folsom}, {H{\'e}brard}, {Jeffers}, {Marsden}, {Morin},
  {Petit}, {Waite}, \& {BCool Collaboration}}]{See2017}
---. 2017, \href{http://dx.doi.org/10.1093/mnras/stw3094}{\JournalTitle{MNRAS},
  466, 1542}

\bibitem[{{Semel}(1989)}]{Semel1989}
{Semel}, M. 1989, \JournalTitle{A\&A}, 225, 456

\bibitem[{{Shulyak} {et~al.}(2017){Shulyak}, {Reiners}, {Engeln}, {Malo},
  {Yadav}, {Morin}, \& {Kochukhov}}]{Shulyak2017}
{Shulyak}, D., {Reiners}, A., {Engeln}, A., {et~al.} 2017,
  \href{http://dx.doi.org/10.1038/s41550-017-0184}{\JournalTitle{Nature
  Astronomy}, 1, 0184}

\bibitem[{{Shulyak} {et~al.}(2014){Shulyak}, {Reiners}, {Seemann}, {Kochukhov},
  \& {Piskunov}}]{Shulyak2014}
{Shulyak}, D., {Reiners}, A., {Seemann}, U., {Kochukhov}, O., \& {Piskunov}, N.
  2014,
  \href{http://dx.doi.org/10.1051/0004-6361/201322136}{\JournalTitle{A\&A},
  563, A35}

\bibitem[{{Stelzer} {et~al.}(2016){Stelzer}, {Damasso}, {Scholz}, \&
  {Matt}}]{Stelzer2016}
{Stelzer}, B., {Damasso}, M., {Scholz}, A., \& {Matt}, S.~P. 2016,
  \href{http://dx.doi.org/10.1093/mnras/stw1936}{\JournalTitle{MNRAS}, 463,
  1844}

\bibitem[{{Strassmeier}(2009)}]{Strassmeier2009}
{Strassmeier}, K.~G. 2009,
  \href{http://dx.doi.org/10.1007/s00159-009-0020-6}{\JournalTitle{A\&AR}, 17,
  251}

\bibitem[{{Suzuki}(2006)}]{Suzuki2006}
{Suzuki}, T.~K. 2006,
  \href{http://dx.doi.org/10.1086/503102}{\JournalTitle{ApJl}, 640, L75}

\bibitem[{{Takeda} {et~al.}(2007){Takeda}, {Ford}, {Sills}, {Rasio}, {Fischer},
  \& {Valenti}}]{Takeda2007}
{Takeda}, G., {Ford}, E.~B., {Sills}, A., {et~al.} 2007,
  \href{http://dx.doi.org/10.1086/509763}{\JournalTitle{ApJS}, 168, 297}

\bibitem[{{Valenti} \& {Fischer}(2005)}]{Valenti2005}
{Valenti}, J.~A., \& {Fischer}, D.~A. 2005,
  \href{http://dx.doi.org/10.1086/430500}{\JournalTitle{ApJS}, 159, 141}

\bibitem[{{Vidotto}(2016)}]{Vidotto2016}
{Vidotto}, A.~A. 2016,
  \href{http://dx.doi.org/10.1093/mnras/stw758}{\JournalTitle{MNRAS}, 459,
  1533}

\bibitem[{{Vidotto} {et~al.}(2018){Vidotto}, {Lehmann}, {Jardine}, \&
  {Pevtsov}}]{Vidotto2018}
{Vidotto}, A.~A., {Lehmann}, L.~T., {Jardine}, M., \& {Pevtsov}, A.~A. 2018,
  \href{http://dx.doi.org/10.1093/mnras/sty1926}{\JournalTitle{MNRAS}, 480,
  477}

\bibitem[{{Vidotto} {et~al.}(2014){Vidotto}, {Gregory}, {Jardine}, {Donati},
  {Petit}, {Morin}, {Folsom}, {Bouvier}, {Cameron}, {Hussain}, {Marsden},
  {Waite}, {Fares}, {Jeffers}, \& {do Nascimento}}]{Vidotto2014Trends}
{Vidotto}, A.~A., {Gregory}, S.~G., {Jardine}, M., {et~al.} 2014,
  \href{http://dx.doi.org/10.1093/mnras/stu728}{\JournalTitle{MNRAS}, 441,
  2361}

\bibitem[{{Waite} {et~al.}(2015){Waite}, {Marsden}, {Carter}, {Petit},
  {Donati}, {Jeffers}, \& {Boro Saikia}}]{Waite2015}
{Waite}, I.~A., {Marsden}, S.~C., {Carter}, B.~D., {et~al.} 2015,
  \href{http://dx.doi.org/10.1093/mnras/stv006}{\JournalTitle{MNRAS}, 449, 8}

\bibitem[{{Waite} {et~al.}(2017){Waite}, {Marsden}, {Carter}, {Petit},
  {Jeffers}, {Morin}, {Vidotto}, {Donati}, \& {BCool
  Collaboration}}]{Waite2017}
---. 2017, \href{http://dx.doi.org/10.1093/mnras/stw2731}{\JournalTitle{MNRAS},
  465, 2076}

\bibitem[{{Wang} \& {Sheeley}(1990)}]{Wang1990}
{Wang}, Y.-M., \& {Sheeley}, Jr., N.~R. 1990,
  \href{http://dx.doi.org/10.1086/168805}{\JournalTitle{ApJ}, 355, 726}

\bibitem[{{Wright} \& {Drake}(2016)}]{Wright2016}
{Wright}, N.~J., \& {Drake}, J.~J. 2016,
  \href{http://dx.doi.org/10.1038/nature18638}{\JournalTitle{Nature}, 535, 526}

\bibitem[{{Wright} {et~al.}(2011){Wright}, {Drake}, {Mamajek}, \&
  {Henry}}]{Wright2011}
{Wright}, N.~J., {Drake}, J.~J., {Mamajek}, E.~E., \& {Henry}, G.~W. 2011,
  \href{http://dx.doi.org/10.1088/0004-637X/743/1/48}{\JournalTitle{ApJ}, 743,
  48}

\bibitem[{{Wright} {et~al.}(2018){Wright}, {Newton}, {Williams}, {Drake}, \&
  {Yadav}}]{Wright2018}
{Wright}, N.~J., {Newton}, E.~R., {Williams}, P.~K.~G., {Drake}, J.~J., \&
  {Yadav}, R.~K. 2018,
  \href{http://dx.doi.org/10.1093/mnras/sty1670}{\JournalTitle{MNRAS}, 479,
  2351}

\bibitem[{{Yadav} {et~al.}(2015){Yadav}, {Christensen}, {Morin}, {Gastine},
  {Reiners}, {Poppenhaeger}, \& {Wolk}}]{Yadav2015}
{Yadav}, R.~K., {Christensen}, U.~R., {Morin}, J., {et~al.} 2015,
  \href{http://dx.doi.org/10.1088/2041-8205/813/2/L31}{\JournalTitle{ApJl},
  813, L31}

\bibitem[{{Yang} {et~al.}(2008){Yang}, {Johns-Krull}, \& {Valenti}}]{Yang2008}
{Yang}, H., {Johns-Krull}, C.~M., \& {Valenti}, J.~A. 2008,
  \href{http://dx.doi.org/10.1088/0004-6256/136/6/2286}{\JournalTitle{AJ}, 136,
  2286}

\end{thebibliography}

\appendix

\begin{table*}
\caption{Stellar parameters for our ZDI sample. Listed are the stellar mass, radius, luminosity, rotation period, Rossby number, average field strength from ZDI, estimated filling factors (see section \ref{subsec:FillingFactors}) and the original publication of the ZDI map. Unless otherwise noted, stellar parameters were taken from the original ZDI publication, \citet{Valenti2005}, \citet{Takeda2007} or \citet{Vidotto2014Trends} and reference therein.}
\label{tab:Params}
\center
\begin{tabular}{lcccccccc}
\hline\hline
Star & $M_{\star}$ & $r_{\star}$ & $L_{\star}$ & $P_{\rm rot}$ & Ro & $\langle B_V\rangle$ & $f_{\rm est}$ & Reference \\
ID & ($M_{\odot}$) & ($r_{\odot}$) & ($L_{\odot}$) & (d) & & (G) & & \\
\hline
											
HD 3651	&	0.88	&	0.88	&	0.52	&	43.4	&	2.1	&	3.58	&	0.085	&	Petit et al. (in prep)	\\
HD 9986	&	1.02	&	1.04	&	1.1	&	23	&	1.8	&	0.605	&	0.029	&	Petit et al. (in prep)	\\
HD 10476	&	0.82	&	0.82	&	0.43	&	16	&	0.74	&	1.98	&	0.055	&	Petit et al. (in prep)	\\
$\kappa$ Cet	&	1.03	&	0.95	&	0.83	&	9.3	&	0.62	&	23.6	&	0.34	&	\citet{Nasciemento2016}	\\
$\epsilon$ Eri (2007)	&	0.86	&	0.74	&	0.33	&	10.3	&	0.45	&	11.8	&	0.18	&	\citet{Jeffers2014}	\\
$\epsilon$ Eri (2008)	&	0.86	&	0.74	&	0.33	&	10.3	&	0.45	&	9.5	&	0.15	&	\citet{Jeffers2014}	\\
$\epsilon$ Eri (2010)	&	0.86	&	0.74	&	0.33	&	10.3	&	0.45	&	15.6	&	0.22	&	\citet{Jeffers2014}	\\
$\epsilon$ Eri (2011)	&	0.86	&	0.74	&	0.33	&	10.3	&	0.45	&	9.84	&	0.16	&	\citet{Jeffers2014}	\\
$\epsilon$ Eri (2012)	&	0.86	&	0.74	&	0.33	&	10.3	&	0.45	&	18.3	&	0.24	&	\citet{Jeffers2014}	\\
$\epsilon$ Eri (2013)	&	0.86	&	0.74	&	0.33	&	10.3	&	0.45	&	19.5	&	0.25	&	\citet{Jeffers2014}	\\
HD 39587	&	1.03	&	1.05	&	1.1	&	4.83	&	0.38	&	18.5	&	0.32	&	Petit et al. (in prep)	\\
HD 56124	&	1.03	&	1.01	&	1.1	&	18	&	1.5	&	2.19	&	0.07	&	Petit et al. (in prep)	\\
HD 72905	&	1	&	1	&	1.1	&	5	&	0.44	&	27.7	&	0.42	&	Petit et al. (in prep)	\\
HD 73350	&	1.04	&	0.98	&	0.95	&	12.3	&	0.93	&	11	&	0.21	&	\citet{Petit2008}	\\
HD 75332	&	1.21	&	1.24	&	2.1	&	4.8	&	0.99	&	6.2	&	0.18	&	Petit et al. (in prep)	\\
HD 76151	&	1.06	&	1	&	0.97	&	20.5	&	1.5	&	2.99	&	0.083	&	\citet{Petit2008}	\\
HD 78366	&	1.13	&	1.06	&	1.2	&	11.4	&	1.1	&	12.3	&	0.24	&	Petit et al. (in prep)	\\
HD 101501	&	0.85	&	0.9	&	0.61	&	17.6	&	0.94	&	12.4	&	0.21	&	Petit et al. (in prep)	\\
$\xi$ Boo A (2007)	&	0.93	&	0.84	&	0.52	&	6.4	&	0.33	&	61.8	&	0.6	&	\citet{Morgenthaler2012}	\\
$\xi$ Boo A (2008)	&	0.93	&	0.84	&	0.52	&	6.4	&	0.33	&	22.2	&	0.29	&	\citet{Morgenthaler2012}	\\
$\xi$ Boo A (2009)	&	0.93	&	0.84	&	0.52	&	6.4	&	0.33	&	36.5	&	0.42	&	\citet{Morgenthaler2012}	\\
$\xi$ Boo A (Jan 2010)	&	0.93	&	0.84	&	0.52	&	6.4	&	0.33	&	29.3	&	0.36	&	\citet{Morgenthaler2012}	\\
$\xi$ Boo A (Jun 2010)	&	0.93	&	0.84	&	0.52	&	6.4	&	0.33	&	24.3	&	0.31	&	\citet{Morgenthaler2012}	\\
$\xi$ Boo A (Jul 2010)	&	0.93	&	0.84	&	0.52	&	6.4	&	0.33	&	35.6	&	0.41	&	\citet{Morgenthaler2012}	\\
$\xi$ Boo A (2011)	&	0.93	&	0.84	&	0.52	&	6.4	&	0.33	&	37.9	&	0.43	&	\citet{Morgenthaler2012}	\\
$\xi$ Boo B	&	0.7$^a$	&	0.55$^b$	&	0.097$^a$	&	10.3	&	0.3	&	16.3	&	0.18	&	Petit et al. (in prep)	\\
18 Sco	&	1.01	&	1.04	&	1.1	&	22.7	&	1.7	&	1.18	&	0.045	&	\citet{Petit2008}	\\
HD 166435	&	1.04	&	0.99	&	0.99	&	3.43	&	0.27	&	20	&	0.32	&	Petit et al. (in prep)	\\
HD 175726	&	1.06	&	1.06	&	1.2	&	3.92	&	0.38	&	9.62	&	0.21	&	Petit et al. (in prep)	\\
HD 190771	&	1.06	&	1.01	&	0.99	&	8.8	&	0.65	&	13.9	&	0.25	&	\citet{Petit2008}	\\
61 Cyg A (2007)	&	0.66	&	0.62	&	0.15	&	34.2	&	1.1	&	11.9	&	0.16	&	\citet{Saikia2016}	\\
61 Cyg A (2008)	&	0.66	&	0.62	&	0.15	&	34.2	&	1.1	&	2.99	&	0.062	&	\citet{Saikia2016}	\\
61 Cyg A (2010)	&	0.66	&	0.62	&	0.15	&	34.2	&	1.1	&	5.49	&	0.096	&	\citet{Saikia2016}	\\
61 Cyg A (2013)	&	0.66	&	0.62	&	0.15	&	34.2	&	1.1	&	9.31	&	0.14	&	\citet{Saikia2016}	\\
61 Cyg A (2014)	&	0.66	&	0.62	&	0.15	&	34.2	&	1.1	&	8.17	&	0.13	&	\citet{Saikia2016}	\\
61 Cyg A (Aug 2015)	&	0.66	&	0.62	&	0.15	&	34.2	&	1.1	&	11.7	&	0.16	&	\citet{Saikia2016}	\\
61 Cyg A (Oct 2015)	&	0.66	&	0.62	&	0.15	&	34.2	&	1.1	&	8.56	&	0.13	&	\citet{Saikia2018}	\\
61 Cyg A (Dec 2015)	&	0.66	&	0.62	&	0.15	&	34.2	&	1.1	&	6.42	&	0.11	&	\citet{Saikia2018}	\\
61 Cyg A (2016)	&	0.66	&	0.62	&	0.15	&	34.2	&	1.1	&	9.08	&	0.14	&	\citet{Saikia2018}	\\
61 Cyg A (Jul 2017)	&	0.66	&	0.62	&	0.15	&	34.2	&	1.1	&	6.69	&	0.11	&	\citet{Saikia2018}	\\
61 Cyg A (Dec 2017)	&	0.66	&	0.62	&	0.15	&	34.2	&	1.1	&	4.35	&	0.081	&	\citet{Saikia2018}	\\
61 Cyg A (2018)	&	0.66	&	0.62	&	0.15	&	34.2	&	1.1	&	9.5	&	0.14	&	\citet{Saikia2018}	\\
HN Peg (2007)	&	1.1	&	1.04	&	1.2	&	4.55	&	0.41	&	18.3	&	0.32	&	\citet{Saikia2015}	\\
HN Peg (2008)	&	1.1	&	1.04	&	1.2	&	4.55	&	0.41	&	14.1	&	0.26	&	\citet{Saikia2015}	\\
HN Peg (2009)	&	1.1	&	1.04	&	1.2	&	4.55	&	0.41	&	11.5	&	0.23	&	\citet{Saikia2015}	\\
HN Peg (2010)	&	1.1	&	1.04	&	1.2	&	4.55	&	0.41	&	19.4	&	0.33	&	\citet{Saikia2015}	\\
HN Peg (2011)	&	1.1	&	1.04	&	1.2	&	4.55	&	0.41	&	19.3	&	0.33	&	\citet{Saikia2015}	\\
HN Peg (2013)	&	1.1	&	1.04	&	1.2	&	4.55	&	0.41	&	23.7	&	0.38	&	\citet{Saikia2015}	\\
HD 219134	&	0.81$^c$	&	0.78$^c$	&	0.27$^c$	&	42.2	&	1.5	&	2.47	&	0.06	&	\citet{Folsom2018Single}	\\
AV 1693	&	0.9	&	0.83	&	0.52	&	9.05	&	0.48	&	33.7	&	0.4	&	\citet{Folsom2018}	\\
AV 1826	&	0.85	&	0.8	&	0.39	&	9.34	&	0.42	&	25.1	&	0.32	&	\citet{Folsom2018}	\\
AV 2177	&	0.9	&	0.78	&	0.43	&	8.98	&	0.45	&	10.3	&	0.17	&	\citet{Folsom2018}	\\
AV 523	&	0.8	&	0.72	&	0.24	&	11.1	&	0.41	&	22.8	&	0.28	&	\citet{Folsom2018}	\\
EP Eri	&	0.85	&	0.72	&	0.3	&	6.76	&	0.29	&	34.3	&	0.37	&	\citet{Folsom2018}	\\
HH Leo	&	0.95	&	0.84	&	0.54	&	5.92	&	0.32	&	28.9	&	0.35	&	\citet{Folsom2018}	\\
Mel25-151	&	0.85	&	0.82	&	0.35	&	10.4	&	0.41	&	23.7	&	0.31	&	\citet{Folsom2018}	\\
Mel25-179	&	0.85	&	0.84	&	0.4	&	9.7	&	0.41	&	26	&	0.33	&	\citet{Folsom2018}	\\
Mel25-21	&	0.9	&	0.91	&	0.56	&	9.73	&	0.47	&	12.6	&	0.21	&	\citet{Folsom2018}	\\
Mel25-43	&	0.85	&	0.79	&	0.38	&	9.9	&	0.44	&	8.52	&	0.15	&	\citet{Folsom2018}	\\
Mel25-5	&	0.85	&	0.91	&	0.43	&	10.6	&	0.42	&	13	&	0.21	&	\citet{Folsom2018}	\\
TYC 1987-509-1	&	0.9	&	0.83	&	0.52	&	9.43	&	0.5	&	25	&	0.32	&	\citet{Folsom2018}	\\
\hline
\end{tabular}
\end{table*}

\begin{table*}
\contcaption{continued}
\center
\begin{tabular}{lcccccccc}
\hline\hline
Star & $M_{\star}$ & $r_{\star}$ & $L_{\star}$ & $P_{\rm rot}$ & Ro & $\langle B_V\rangle$ & $f_{\rm est}$ & Reference \\
ID & ($M_{\odot}$) & ($r_{\odot}$) & ($L_{\odot}$) & (d) & & (G) & & \\
\hline		
V447 Lac	&	0.9	&	0.81	&	0.46	&	4.43	&	0.22	&	39	&	0.43	&	\citet{Folsom2016}	\\
DX Leo	&	0.9	&	0.81	&	0.49	&	5.38	&	0.28	&	29.1	&	0.35	&	\citet{Folsom2016}	\\
V439 And	&	0.95	&	0.92	&	0.64	&	6.23	&	0.33	&	13.9	&	0.22	&	\citet{Folsom2016}	\\
																	
\textbf{Young Suns} \\																	
AB Dor (2001)	&	0.9$^d$	&	0.96$^e$	&	0.63$^f$	&	0.51	&	0.025	&	239	&	1.7	&	\citet{Donati2003}	\\
AB Dor (2002)	&	0.9$^d$	&	0.96$^e$	&	0.63$^f$	&	0.51	&	0.025	&	198	&	1.5	&	\citet{Donati2003}	\\
BD-16351	&	0.9	&	0.88	&	0.52	&	3.21	&	0.15	&	49	&	0.53	&	\citet{Folsom2016}	\\
HII 296	&	0.9	&	0.93	&	0.49	&	2.61	&	0.11	&	80.4	&	0.77	&	\citet{Folsom2016}	\\
HII 739	&	1.15	&	1.07	&	1.4	&	1.58	&	0.18	&	15.4	&	0.29	&	\citet{Folsom2016}	\\
HIP 12545	&	0.95	&	1.07	&	0.4	&	4.83	&	0.14	&	116	&	0.97	&	\citet{Folsom2016}	\\
HIP 76768	&	0.8	&	0.85	&	0.27	&	3.7	&	0.12	&	113	&	0.9	&	\citet{Folsom2016}	\\
Lo Peg	&	0.75	&	0.66	&	0.2	&	0.423	&	0.015	&	140	&	0.96	&	\citet{Folsom2016}	\\
PELS 031	&	0.95	&	1.05	&	0.62	&	2.5	&	0.1	&	44.1	&	0.53	&	\citet{Folsom2016}	\\
PW And	&	0.85	&	0.78	&	0.35	&	1.76	&	0.075	&	126	&	0.97	&	\citet{Folsom2016}	\\
TYC 0486-4943-1	&	0.75	&	0.69	&	0.21	&	3.75	&	0.13	&	25	&	0.29	&	\citet{Folsom2016}	\\
TYC 5164-567-1	&	0.9	&	0.89	&	0.5	&	4.68	&	0.21	&	63.9	&	0.64	&	\citet{Folsom2016}	\\
TYC 6349-0200-1	&	0.85	&	0.96	&	0.3	&	3.41	&	0.1	&	59.7	&	0.58	&	\citet{Folsom2016}	\\
TYC 6878-0195-1	&	1.17	&	1.37	&	0.8	&	5.7	&	0.18	&	55.3	&	0.66	&	\citet{Folsom2016}	\\
HD 6569	&	0.85	&	0.76	&	0.36	&	7.13	&	0.32	&	25	&	0.31	&	\citet{Folsom2018}	\\
HIP 10272	&	0.9	&	0.8	&	0.45	&	6.13	&	0.31	&	21.2	&	0.28	&	\citet{Folsom2018}	\\
BD-072388	&	0.85	&	0.78	&	0.38	&	0.326	&	0.015	&	195	&	1.3	&	\citet{Folsom2018}	\\
HD 141943 (2007)	&	1.3	&	1.6	&	2.8	&	2.18	&	0.18	&	92.7	&	1.3	&	\citet{Marsden2011}	\\
HD 141943 (2009)	&	1.3	&	1.6	&	2.8	&	2.18	&	0.18	&	37.3	&	0.66	&	\citet{Marsden2011}	\\
HD 141943 (2010)	&	1.3	&	1.6	&	2.8	&	2.18	&	0.18	&	71.7	&	1.1	&	\citet{Marsden2011}	\\
HD 35296 (2007)	&	1.06	&	1.1	&	1.6$^g$	&	3.48	&	0.56	&	13.5	&	0.3	&	\citet{Waite2015}	\\
HD 35296 (2008)	&	1.06	&	1.1	&	1.6$^g$	&	3.48	&	0.56	&	18.1	&	0.36	&	\citet{Waite2015}	\\
HD 29615	&	0.95	&	1	&	1$^h$	&	2.34	&	0.19	&	85.6	&	0.94	&	\citet{Waite2015}	\\
EK Dra (2006)	&	0.95	&	0.94	&	0.76$^i$	&	2.77	&	0.17	&	92.9	&	0.89	&	\citet{Waite2017}	\\
EK Dra (Jan 2007)	&	0.95	&	0.94	&	0.76$^i$	&	2.77	&	0.17	&	73.8	&	0.76	&	\citet{Waite2017}	\\
EK Dra (Feb 2007)	&	0.95	&	0.94	&	0.76$^i$	&	2.77	&	0.17	&	52	&	0.59	&	\citet{Waite2017}	\\
EK Dra (2008)	&	0.95	&	0.94	&	0.76$^i$	&	2.77	&	0.17	&	54.8	&	0.62	&	\citet{Waite2017}	\\
EK Dra (2012)	&	0.95	&	0.94	&	0.76$^i$	&	2.77	&	0.17	&	96.4	&	0.92	&	\citet{Waite2017}	\\
																	
\textbf{Hot Jupiter Hosts} \\																	
$\tau$ Boo (Jan 2008)	&	1.34	&	1.46	&	3	&	3	&	0.66	&	2.46	&	0.11	&	\citet{Fares2009}	\\
$\tau$ Boo (Jun 08)	&	1.34	&	1.46	&	3	&	3	&	0.66	&	1.52	&	0.075	&	\citet{Fares2009}	\\
$\tau$ Boo (Jul 2008)	&	1.34	&	1.46	&	3	&	3	&	0.66	&	1.27	&	0.066	&	\citet{Fares2009}	\\
$\tau$ Boo (2009)	&	1.34	&	1.46	&	3	&	3	&	0.66	&	1.99	&	0.091	&	\citet{Fares2013}	\\
$\tau$ Boo (2010)	&	1.34	&	1.46	&	3	&	3	&	0.66	&	2.94	&	0.12	&	\citet{Fares2013}	\\
$\tau$ Boo (Jan 2011)	&	1.34	&	1.46	&	3	&	3	&	0.66	&	2.58	&	0.11	&	\citet{Fares2013}	\\
$\tau$ Boo (May 2011)	&	1.34	&	1.46	&	3	&	3	&	0.66	&	2.47	&	0.11	&	\citet{Mengel2016}	\\
$\tau$ Boo (May 2013)	&	1.34	&	1.46	&	3	&	3	&	0.66	&	2.45	&	0.1	&	\citet{Mengel2016}	\\
$\tau$ Boo (Dec 2013)	&	1.34	&	1.46	&	3	&	3	&	0.66	&	3.85	&	0.14	&	\citet{Mengel2016}	\\
$\tau$ Boo (2014)	&	1.34	&	1.46	&	3	&	3	&	0.66	&	1.82	&	0.085	&	\citet{Mengel2016}	\\
$\tau$ Boo (Jan 2015)	&	1.34	&	1.46	&	3	&	3	&	0.66	&	2.54	&	0.11	&	\citet{Mengel2016}	\\
$\tau$ Boo (2 Apr 2015)	&	1.34	&	1.46	&	3	&	3	&	0.66	&	1.18	&	0.063	&	\citet{Mengel2016}	\\
$\tau$ Boo (13 Apr 2015)	&	1.34	&	1.46	&	3	&	3	&	0.66	&	0.905	&	0.052	&	\citet{Mengel2016}	\\
$\tau$ Boo (20 Apr 2015)	&	1.34	&	1.46	&	3	&	3	&	0.66	&	1.19	&	0.063	&	\citet{Mengel2016}	\\
$\tau$ Boo (May 2015)	&	1.34	&	1.46	&	3	&	3	&	0.66	&	1.95	&	0.089	&	\citet{Mengel2016}	\\
HD 73256	&	1.05	&	0.89	&	0.72	&	14	&	0.93	&	6.2	&	0.13	&	\citet{Fares2013}	\\
HD 102195	&	0.87	&	0.82	&	0.48	&	12.3	&	0.62	&	10.7	&	0.18	&	\citet{Fares2013}	\\
HD 130322	&	0.79	&	0.83	&	0.5	&	26.1	&	1.3	&	2.34	&	0.063	&	\citet{Fares2013}	\\
HD 179949 (2007)	&	1.21	&	1.19	&	1.8	&	7.6	&	1.2	&	2.29	&	0.086	&	\citet{Fares2012}	\\
HD 179949 (2009)	&	1.21	&	1.19	&	1.8	&	7.6	&	1.2	&	3.17	&	0.11	&	\citet{Fares2012}	\\
HD 189733 (2007)	&	0.82	&	0.76	&	0.34	&	12.5	&	0.54	&	19.6	&	0.26	&	\citet{Fares2010}	\\
HD 189733 (2008)	&	0.82	&	0.76	&	0.34	&	12.5	&	0.54	&	32.4	&	0.37	&	\citet{Fares2010}	\\
																	
\textbf{M dwarf Stars} \\																	
CE Boo	&	0.48	&	0.43	&	0.033	&	14.7	&	0.32	&	103	&	0.51	&	\citet{Donati2008}	\\
DS Leo (2007)	&	0.58	&	0.52	&	0.052	&	14	&	0.32	&	101	&	0.54	&	\citet{Donati2008}	\\
DS Leo (2008)	&	0.58	&	0.52	&	0.052	&	14	&	0.32	&	86.9	&	0.49	&	\citet{Donati2008}	\\
GJ 182	&	0.75	&	0.82	&	0.13	&	4.35	&	0.099	&	172	&	0.96	&	\citet{Donati2008}	\\
GJ 49	&	0.57	&	0.51	&	0.052	&	18.6	&	0.43	&	27	&	0.21	&	\citet{Donati2008}	\\
AD Leo (2007)	&	0.42	&	0.38	&	0.021	&	2.24	&	0.044	&	167	&	0.72	&	\citet{Morin2008}	\\
AD Leo (2008)	&	0.42	&	0.38	&	0.021	&	2.24	&	0.044	&	178	&	0.76	&	\citet{Morin2008}	\\
DT Vir (2007)	&	0.59	&	0.53	&	0.055	&	2.85	&	0.065	&	145	&	0.7	&	\citet{Donati2008}	\\
DT Vir (2008)	&	0.59	&	0.53	&	0.055	&	2.85	&	0.065	&	149	&	0.72	&	\citet{Donati2008}	\\
EQ Peg A	&	0.39	&	0.35	&	0.018	&	1.06	&	0.021	&	416	&	1.3	&	\citet{Morin2008}	\\
EQ Peg B	&	0.25	&	0.25	&	0.0072	&	0.4	&	0.0071	&	414	&	1.4	&	\citet{Morin2008}	\\
\hline
\end{tabular}
\end{table*}

\begin{table*}
\contcaption{continued}
\center
\begin{tabular}{lcccccccc}
\hline\hline
Star & $M_{\star}$ & $r_{\star}$ & $L_{\star}$ & $P_{\rm rot}$ & Ro & $\langle B_V\rangle$ & $f_{\rm est}$ & Reference \\
ID & ($M_{\odot}$) & ($r_{\odot}$) & ($L_{\odot}$) & (d) & & (G) & & \\
\hline		
EV Lac (2006)	&	0.32	&	0.3	&	0.013	&	4.37	&	0.085	&	523	&	1.5	&	\citet{Morin2008}	\\
EV Lac (2007)	&	0.32	&	0.3	&	0.013	&	4.37	&	0.085	&	463	&	1.4	&	\citet{Morin2008}	\\
DX Cnc (2007)	&	0.1	&	0.11	&	0.0006	&	0.46	&	0.0059	&	112	&	0.35	&	\citet{Morin2010}	\\
DX Cnc (2008)	&	0.1	&	0.11	&	0.0006	&	0.46	&	0.0059	&	76.6	&	0.27	&	\citet{Morin2010}	\\
DX Cnc (2009)	&	0.1	&	0.11	&	0.0006	&	0.46	&	0.0059	&	77.1	&	0.27	&	\citet{Morin2010}	\\
GJ 1156 (2007)	&	0.14	&	0.16	&	0.0025	&	0.49	&	0.0081	&	47	&	0.28	&	\citet{Morin2010}	\\
GJ 1156 (2008)	&	0.14	&	0.16	&	0.0025	&	0.49	&	0.0081	&	98.2	&	0.47	&	\citet{Morin2010}	\\
GJ 1156 (2009)	&	0.14	&	0.16	&	0.0025	&	0.49	&	0.0081	&	84.9	&	0.42	&	\citet{Morin2010}	\\
GJ 1245 B (2006)	&	0.12	&	0.14	&	0.0016	&	0.71	&	0.011	&	164	&	0.66	&	\citet{Morin2010}	\\
GJ 1245 B (2008)	&	0.12	&	0.14	&	0.0016	&	0.71	&	0.011	&	55.4	&	0.31	&	\citet{Morin2010}	\\
OT Ser	&	0.55	&	0.49	&	0.041	&	3.4	&	0.073	&	123	&	0.61	&	\citet{Donati2008}	\\
V374 Peg (2005)	&	0.28	&	0.28	&	0.0095	&	0.45	&	0.0082	&	706	&	2	&	\citet{Morin2008V374Peg}	\\
V374 Peg (2006)	&	0.28	&	0.28	&	0.0095	&	0.45	&	0.0082	&	596	&	1.8	&	\citet{Morin2008V374Peg}	\\
WX UMa (2006)	&	0.1	&	0.12	&	0.00081	&	0.78	&	0.01	&	1010	&	1.9	&	\citet{Morin2010}	\\
WX UMa (2007)	&	0.1	&	0.12	&	0.00081	&	0.78	&	0.01	&	1250	&	2.2	&	\citet{Morin2010}	\\
WX UMa (2008)	&	0.1	&	0.12	&	0.00081	&	0.78	&	0.01	&	1240	&	2.2	&	\citet{Morin2010}	\\
WX UMa (2009)	&	0.1	&	0.12	&	0.00081	&	0.78	&	0.01	&	1670	&	2.7	&	\citet{Morin2010}	\\
YZ CMi (2007)	&	0.32	&	0.29	&	0.012	&	2.77	&	0.054	&	579	&	1.6	&	\citet{Morin2008}	\\
YZ CMi (2008)	&	0.32	&	0.29	&	0.012	&	2.77	&	0.054	&	533	&	1.5	&	\citet{Morin2008}	\\
GJ 176	&	0.49	&	0.47	&	0.033	&	39.3	&	0.79	&	30.2	&	0.24	&	H\'{e}brard et al. (in prep)	\\
GJ 205	&	0.63	&	0.55	&	0.061	&	33.6	&	0.78	&	19.6	&	0.17	&	\citet{Hebrard2016}	\\
GJ 358	&	0.42	&	0.41	&	0.023	&	25.4	&	0.49	&	125	&	0.63	&	\citet{Hebrard2016}	\\
GJ 479	&	0.43	&	0.42	&	0.025	&	24	&	0.47	&	58	&	0.37	&	\citet{Hebrard2016}	\\
GJ 674	&	0.35	&	0.4	&	0.016	&	35.2	&	0.59	&	131	&	0.74	&	H\'{e}brard et al. (in prep)	\\
GJ 846 (2013)	&	0.6	&	0.54	&	0.059	&	10.7	&	0.25	&	20.3	&	0.18	&	\citet{Hebrard2016}	\\
GJ 846 (2014)	&	0.6	&	0.54	&	0.059	&	10.7	&	0.25	&	26.9	&	0.22	&	\citet{Hebrard2016}	\\
\hline
\end{tabular}
\small
\item{$^a$ \citet{Fernandes1998}, $^b$ \citet{Cranmer2011}, $^c$ \citet{Gillon2017}, $^d$ \citet{Azulay2017},  $^e$ \citet{Guirado2011},\\
$^f$calculated using Stefan-Boltzmann law with $T_{\rm eff}$ = 5250 K \citep{Strassmeier2009},\\
$^g$calculated using Stefan-Boltzmann law with $T_{\rm eff}$ = 6170 K \citep{Waite2015},\\
$^h$calculated using Stefan-Boltzmann law with $T_{\rm eff}$ = 5820 K \citep{Waite2015},\\
$^i$calculated using Stefan-Boltzmann law with $T_{\rm eff}$ = 5561 K \citep{Waite2017}}

\end{table*}

\begin{table*}
\caption{Magnetic field strengths obtained using the Zeeman broadening technique from the literature. When a field strength, $B$, and a filling factor, $f$, are listed individually in the original paper, both are shown here. Only a single number is listed when the original paper lists a combined $Bf$ value.}
\label{tab:ZB}
\center
\begin{tabular}{lcclcc}
\hline\hline
Star	&	$\langle B_I\rangle$ & Reference \hspace{5mm} & Star	&	$\langle B_I\rangle$ & Reference \\	
ID & (G) & \hspace{5mm} & ID & (G) &\\	
\hline
$\kappa$ Cet	&	321	&	\citet{Montesinos1993}	\hspace{5mm}	&	GJ 1156	&	2100	&	\citet{Reiners2009FluxSat}	\\
... 	&	392	&	\citet{Montesinos1993}	\hspace{5mm}	&	WX Uma	&	7300	&	\citet{Shulyak2017}	\\
... 	&	406	&	\citet{Montesinos1993}	\hspace{5mm}	&	EV Lac	&	3900	&	\citet{Reiners2007}	\\
... 	&	480	&	\citet{Montesinos1993}	\hspace{5mm}	&	... 	&	4200	&	\citet{Shulyak2017}	\\
... 	&	1500$\times$0.35	&	\citet{Montesinos1993}	\hspace{5mm}	&	... 	&	3900	&	\citet{Saar2001}	\\
$\xi$ Boo A	&	1600$\times$0.22	&	\citet{Marcy1989}	\hspace{5mm}	&	YZ Cmi	&	3300	&	\citet{Saar2001}	\\
... 	&	1800$\times$0.35	&	\citet{Montesinos1993}	\hspace{5mm}	&	... 	&	4800	&	\citet{Shulyak2017}	\\
... 	&	2000$\times$0.2	&	\citet{Montesinos1993}	\hspace{5mm}	&	GJ 1245 B	&	1700	&	\citet{Reiners2007}	\\
... 	&	1900$\times$0.18	&	\citet{Cranmer2011}	\hspace{5mm}	&	... 	&	3400	&	\citet{Shulyak2017}	\\
$\xi$ Boo B	&	2300$\times$0.2	&	\citet{Saar1994}	\hspace{5mm}	&	DX Cnc	&	1700	&	\citet{Reiners2007}	\\
$\epsilon$ Eri	&	165	&	\citet{Saar2001}	\hspace{5mm}	&	... 	&	3200	&	\citet{Shulyak2017}	\\
... 	&	1000$\times$0.3	&	\citet{Marcy1989}	\hspace{5mm}	&	CE Boo	&	1750	&	\citet{Reiners2009}	\\
... 	&	1900$\times$0.12	&	\citet{Montesinos1993}	\hspace{5mm}	&	... 	&	1800	&	\citet{Shulyak2017}	\\
... 	&	1440$\times$0.088	&	\citet{Cranmer2011}	\hspace{5mm}	&	GJ 182	&	2730	&	\citet{Reiners2009}	\\
61 Cyg A	&	1200$\times$0.24	&	\citet{Marcy1989}	\hspace{5mm}	&	... 	&	2600	&	\citet{Shulyak2017}	\\
DT Vir	&	3000$\times$0.5	&	\citet{Cranmer2011}	\hspace{5mm}	&	DS Leo	&	900	&	\citet{Shulyak2017}	\\
... 	&	2600	&	\citet{Shulyak2017}	\hspace{5mm}	&	OT Ser	&	2700	&	\citet{Shulyak2017}	\\
AD Leo	&	3300	&	\citet{Saar2001}	\hspace{5mm}	&	GJ 49	&	800	&	\citet{Shulyak2017}	\\
... 	&	4000$\times$0.6	&	\citet{Cranmer2011}	\hspace{5mm}	&	EQ Peg A	&	3600	&	\citet{Shulyak2017}	\\
... 	&	2900	&	\citet{Reiners2007}	\hspace{5mm}	&	EQ Peg B	&	4200	&	\citet{Shulyak2017}	\\
... 	&	3100	&	\citet{Shulyak2017}	\hspace{5mm}	&	V374 Peg	&	5300	&	\citet{Shulyak2017}	\\
\hline
\end{tabular}
\end{table*}

\end{document}